\documentclass[12pt,twoside]{article}   %For printing on two sides of page
\usepackage[super,sort,comma]{natbib}
\usepackage{titlesec}

\titlespacing*{\section}
  {0pt}   % left margin
  {12pt plus 2pt minus 2pt} % space before
  {6pt}   % space after

\titlespacing*{\subsection}
  {0pt}   % left margin
  {12pt plus 2pt minus 2pt} % space before
  {6pt}   % space after

\usepackage{authblk}
\usepackage{fancyhdr}		%Gives headers and footers defined below
\usepackage[section]{placeins}   %
\usepackage{float}
\usepackage[version=4]{mhchem}
\usepackage{ulem}
\usepackage{xcolor} 
\usepackage{graphicx}
\usepackage{stmaryrd}
\usepackage{amsmath}
\usepackage{multirow}
\makeatletter \renewcommand\@biblabel[1]{$^{#1}$} \makeatother
\newcommand{\cen}[1]{\begin{center} #1 \end{center}}

\usepackage[mathlines]{lineno}
\usepackage{hyperref}
\hypersetup{ colorlinks,
    citecolor=blue,
    filecolor=blue,
    linkcolor=blue,
    urlcolor=blue
}
\newcommand{\muu}{\text{\textmu}}

\usepackage{xcolor}
\definecolor{gray}{rgb}{0.6,0.6,0.6}
\definecolor{red}{rgb}{0.85,0,0}
\definecolor{green}{rgb}{0,0.85,0}
\definecolor{blue}{rgb}{0,0,0.85}
\definecolor{beige}{rgb}{0.92,0.87,0.78}
\usepackage[all]{hypcap}    %causes link to figures to go to figure, not caption

\titlespacing*{\section}{0pt}{12pt}{6pt}
\titlespacing*{\subsection}{0pt}{10pt}{6pt}
\titlespacing*{\subsubsection}{0pt}{8pt}{6pt}
\begin{document}
%%%%%%%%%%%%%%%%%%%%%%%%%%%%%%%%%%%%%%%%%%%%%%%%%%%%%%%%%%%%%%%%%%%%%%%%%%%
%%%%%%%%%%%%%%%%%%%%%%%%%%%% FRONT PAGE  %%%%%%%%%%%%%%%%%%%%%%%%%%%%%%%%%%
%%%%%%%%%%%%%%%%%%%%%%%%%%%%%%%%%%%%%%%%%%%%%%%%%%%%%%%%%%%%%%%%%%%%%%%%%%%
\thispagestyle{empty}
\cen{\sf {\huge {\bfseries {Advancements in Monte Carlo simulations with gMicroMC: reactive species build-up promotes radical-radical reactions at Flash dose rates}} \\  
\vspace{0.75cm}

\Large
Miguel Molina-Hernández\textsuperscript{1, 2, 3}, Patrícia Gonçalves\textsuperscript{2, 3}, Yujie Chi\textsuperscript{4, *}, João Seco\textsuperscript{1, 5, $\dagger$}} \\

\vspace{0.25cm}

\textsuperscript{1}German Cancer Research Center, Biomedical Physics in Radiation Oncology, Heidelberg, Germany; \textsuperscript{2}Laboratory of Instrumentation and Experimental Particles Physics, Lisbon, Portugal; \textsuperscript{3}University Técnico Lisboa, Physical Engineering, Lisbon, Portugal; \textsuperscript{4}Physics Department, The University of Texas at Arlington, Arlington, Texas, United States of America;  \textsuperscript{5}University of Heidelberg, Physics and Astronomy, Heidelberg, Germany

\vspace{0.25cm}
\today
}
\vspace{-0.25cm}

% \pagenumbering{roman}
% \setcounter{page}{1}
% \pagestyle{plain}
\textsuperscript{*}Joint Last Author. \textsuperscript{$\dagger$}Author to whom correspondence should be addressed. email: j.seco@heidelberg-dkfz.de Website: https://www.dkfz.de/en/medizinische-physik-radioonkologie/index.php
\vspace{-0.5cm}
\begin{abstract}
\vspace{-0.25cm}
\noindent Ultra-high dose rate irradiations to water indicate an enhancement of radical-radical reactions, which could potentially correlate with the Flash effect. The purpose of this work was to extend gMicroMC to support multiple pulse simulations and Flash dose rates, and to investigate, in a pure water model, the mechanisms underlying the enhancement of radical-radical reactions under Flash conditions. gMicroMC, a GPU-based Monte Carlo track-structure algorithm, was extended to simulate multiple pulses. Pure water was exposed to multiple 70 MeV protons pulses delivering up to 20 Gy. The pulse dose rate was set to $2 \cdot 10^5$ and $10^6$ Gy/s, while the average dose rate ranged from 0.01 to 100000 Gy/s. The G-values of H$_2$O$_2$ were used to monitor the influence of dose rate on radical-radical reactions. The multiple pulse extension of gMicroMC was validated against Kinetiscope. Multiple pulse simulations indicated an average dose rate threshold. Below it, complete radical depletion occurred within the pulses, leading to constant G-values. Above it, reactive species accumulated throughout the irradiation, resulting in an increase of radical-radical reactions and thus the G-values of H$_2$O$_2$. The average dose rate thresholds were in the order of 10 and 100 Gy/s for pulse dose rates of $2 \cdot 10^5$ and $10^6$ Gy/s, respectively. At ultra-high dose rates, the brief intervals between pulses led to a reactive species build-up, which enhanced radical-radical reactions. This build-up is more likely to promote radical-radical reactions than the inter-track mechanism. The advancements in gMicroMC provide a sophisticated tool to study chemical dose rate dependencies.\\
% \textbf{Keywords:} Radiolysis, Monte Carlo track-structure, G-values, reactive species build-up, H$_2$O$_2$.\\\\
\end{abstract}

% \begin{table}[h]
%   \begin{tabular}{ll}    
%     \textbf{Abbreviations} & \\          
%     \textbf{G$\boldsymbol{_0}$} & Primary G-values \\        
%     \textbf{MCTS} & Monte Carlo track-structure \\    
%     \textbf{NumODEs} & Numerical ordinary differential equations \\  
%     \textbf{PBCs} & Periodic boundary conditions \\    
%     \textbf{SBS} & Step-by-step \\
%     \textbf{SDR} & Standard dose rate \\    
%     \textbf{UHDR} & Ultra-high dose rate \\
%     \textbf{VOI} & Volume of interest \\
%   \end{tabular}%  
% \end{table}

%\newpage     %may or may not be needed

%\tableofcontents

%\newpage

%%%%%%%%%%%%%%%%%%%%%%%%%%%%%%%%%%%%%%%%%%%%%%%%%%%%%%%%%%%%%%
%%%%%%%%%%%%%%%%%%%%%%%%%%%%%%%%%%%%%%%%%%%%%%%%%%%%%%%%%%%%%%
%%%%%%%%%%%%%%%%%%%%%%%%%%%%%%%%%%%%%%%%%%%%%%%%%%%%%%%%%%%%%%

\section{Introduction}
\setlength{\baselineskip}{0.7cm}      %double spacing		

\pagenumbering{arabic}
\setcounter{page}{1}
\pagestyle{fancy}

\label{introduction}
In recent years, irradiations with ultra-high dose rate (UHDR), above 40 Gy/s, have garnered attention due to the increased differential response between normal tissue and cancer \cite{favaudon2014ultrahigh}. The Flash effect\cite{montay2017irradiation, vozenin2019biological, durante2018faster} is characterized by the radioprotection of the healthy tissue while preserving the clinical tumor control achieved with standard dose rate (SDR). Ever since the advent of this novel technique, it has been discussed what might be the mechanism that could be triggering the Flash effect \cite{vozenin2022towards}. In this work, the chemical species generated through the radiolysis of water were studied as a possible surrogate of the Flash effect.

In the context of radiotherapy, radiation chemistry explores the chemical changes induced by radiation interacting with the cell environment, primarily involving water molecules. A fundamental aspect of this field is water’s radiolysis, which consists of the rupture of water molecules by radiation and the subsequent creation of chemical species (radicals and molecules) which provoke damage to the DNA. The radiolysis of water is traditionally described as the process initiated by an independent primary radiation, encompassing the physical, physico-chemical, and chemical stages. Firstly, during the physical stage (10$^{-17}$ - 10$^{-15}$ s), the radiation excites and ionizes water molecules, leading to the ejection of electrons. Following this, the physico-chemical stage (10$^{-15}$ – 10$^{-12}$ s) takes place, which consists of the dissociation and decay of the excited and ionized water molecules into the first chemical products, plus the thermalization of the electrons. The next stage is the non-homogeneous chemical stage, which begins approximately at 10$^{-12}$ s. During this phase, the chemicals are distributed along the radiation’s pathway, often referred to as the chemical track, and react among each other as they diffuse through Brownian motion. 

The radiolysis process has been extensively investigated by examining the primary chemical yield through pulse radiolysis experiments and scavenging techniques\cite{elliot1994rate}. The G-values represents the number of chemical species generated by radiation normalized to the deposited energy, typically presented either as 10$^{-7}$·mol/J or Species/100 eV. G$_0$ specifically denotes the G-values by an independent chemical track measured at a time less than 1 $\muu$s since its creation. It is often referred as the escape yields of chemical species, and it is characterized by measuring the equilibrium yield after irradiation. Paralelly, Monte Carlo track-structure (MCTS) algorithms have characterized the G$_0$ by simulating independent tracks up to 1 $\muu$s. Among these MCTS models are included gMicroMC \cite{tian2017accelerated, tsai2020new}, TRAX-Chem \cite{boscolo2018trax}, TOPAS-nBio \cite{ramos2018monte}, Geant4-DNA \cite{incerti2010comparison}, PARTRAC \cite{ballarini2000stochastic} and RITRACKS \cite{plante2011monte, plante2013monte}. Experimental measurements in conjunction with MCTS modeling have enabled the characterization of the G$_0$. Nevertheless, the chemical yield after a multiple pulse irradiation is not as straightforward to simulate since numerous tracks synchronize in time and space.

In the latest years, comprehensive studies of the macroscopic chemical yield were conducted to explore potential dose rate dependencies. One key area of debate is the G-value of H$_2$O$_2$ post-irradiation. While traditional pulse radiolysis predict an increase with dose rate \cite{wardman2020radiotherapy}, newer studies have observed a decrease \cite{montay2019long, kacem2022comparing, blain2022proton, sunnerberg2023mean}. On the other hand, it was shown experimentally  that the oxygen consumption rate decreases under UHDR \cite{jansen2021does, cao2021quantification, el2022ultrafast, sunnerberg2023mean}. The experimentalists argued that UHDR could promote reactions such as the recombination of $e_{\text{aq}}^-$ with $^{\cdot}$OH, thus surrogating the H$_2$O$_2$ reduction, or the reactions of $e_{\text{aq}}^-$ and $^{\cdot}$H with O$_2$, which could account for the decrease of oxygen consumption. Concurrently, to corroborate these experimental dose rate dependencies, the capabilities of MCTS algorithms were improved. Several research groups modeled multiple simultaneous tracks up to 1 $\muu$s to study inter-track chemical reactions as a possible surrogate for the G$_0$ dose rate dependencies \cite{kreipl2009time, alanazi2021computer, abolfath2022effect, thompson2023investigating, derksen2023method}. The inter-track hypothesis proposes that under UHDR, two or more chemical tracks might coincide in time and space, allowing radicals from different tracks to interact, thus facilitating radical-radical reactions. However, several studies \cite{weber2022flash, thompson2023investigating} reveal that at clinically relevant dose levels, inter-track coincidences occur infrequently, making inter-track hypothesis doubtful. Meanwhile, other more sophisticated works incorporated the time delay between the tracks, effectively simulating a pulse beam \cite{ramos2020let, lai2021modeling}. Additionally, some researchers introduced new techniques to trace the species beyond 1 $\muu$s and during the homogeneous stage \cite{tran2021geant4, d2023integrated, camazzola2023trax}.

In this work, we use gMicroMC\cite{tian2017accelerated}, a GPU-based MCTS algorithm designed to understand the effects of radiation at the cellular and sub-cellular level. It integrates physical, physico-chemical, and chemical modules adopted from PARTRAC and Geant4-DNA\cite{tsai2020new, tsai2020performance}, utilizing the step-by-step (SBS) method for the chemical’s transportation, being all reactions diffusion-controlled\cite{tian2017accelerated}. It is also used for the simulation of DNA damage\cite{lai2021recent}. Further, Lai et al.\cite{lai2021modeling} demonstrated with pulse simulations an oxygen consumption rate reduction between 0.22 $\muu$M/Gy and 0.19 $\muu$M/Gy as pulse dose rate ($\dot{D}_p$) escalates from 10$^6$ to 10$^8$ Gy/s, matching the experimental tendency. In this study, we extended gMicroMC to simulate multiple pulse irradiations. It integrates periodic boundary conditions (PBCs) to represent the macroscopic properties of water and employs numerical ordinary differential equations (NumODEs) to accelerate the calculation of the homogeneous stage. We reproduced single pulse radiolysis experiments to validate the new techniques. Additionally, we simulated multiple pulse radiolysis for SDR and UHDR conditions using exclusively 70 MeV protons, which have a low LET below 1 keV/$\muu$m. Both the single and multiple pulse radiolysis results were compared against experimental data. 

Our results indicate that while long-lived chemical species such as \ce{H2} and \ce{H2O2} accumulate under both SDR and UHDR conditions, short-lived and highly reactive species such as \ce{e$_{\text{aq}}^-$} and \ce{$^{\cdot}$OH} accumulate exclusively under UHDR conditions. This buildup enhances radical-radical reactions, we refer to  it as the reactive species build-up, and we propose it as an alternative to the inter-track hypothesis, which only applies to particles within the same pulse. Given that our observations involve temporally separated radiation events, this represents a distinct effect. It contributes to understanding how radical-radical reactions are enhanced under UHDR conditions, an enhancement suggested to potentially surrogate the Flash effect\cite{jansen2022changes}. Variances in the radical's yield might affect DNA damage, underscoring the importance of the reactive species build-up.

%%%%%%%%%%%%%%%%%%%%%%%%%%%%%%%%%%%%%%%%%%%%%%%%%%%%%%%%%%%%%%
%%%%%%%%%%%%%%%%%%%%%%%%%%%%%%%%%%%%%%%%%%%%%%%%%%%%%%%%%%%%%%
%%%%%%%%%%%%%%%%%%%%%%%%%%%%%%%%%%%%%%%%%%%%%%%%%%%%%%%%%%%%%%

\section{Methods}
\label{methods}
A pulse irradiation model that operates under the input parameters of accumulated dose, dose rate, and pulse beam structure was developed. It ensures a safe representation of macroscopic irradiations within the microscopic simulated system. To achieve this, gMicroMC’s capabilities were enhanced, moving forward from the usual MCTS modeling of independent tracks to simulations involving multiple tracks. The spatial and temporal proximity of the tracks aligns with the mentioned macroscopic inputs. Additionally, the chemicals are accumulated between pulses and traced beyond the $\muu$s scale into seconds and minutes while ensuring ongoing chemical reactions.

%%%%%%%%%%%%%%%%%%%%%%%%%%%%%%%%%%%%%%%%%%%%%%%%%%%%%%%%%%%%%%
%%%%%%%%%%%%%%%%%%%%%%%  Model WORKFLOW %%%%%%%%%%%%%%%%%%%%%%
%%%%%%%%%%%%%%%%%%%%%%%%%%%%%%%%%%%%%%%%%%%%%%%%%%%%%%%%%%%%%%

\subsection{Workflow of gMicroMC extension}
The model firstly takes as input parameters the targeted dose ($D$) in the micrometric volume of interest (VOI), which is characterized by its side length ($L$), and the particle's kinetic energy ($E$) (Figure \ref{Flowchart} in Supplement). Concerning the pulse beam structure, the input parameters are the $\dot{D}_p$, the average dose rate ($\dot{D}_{av}$), the pulse full width at half maximum (FWHM), and its frequency ($f$). The $\dot{D}_{p}$ is equivalent to $D_p$/FWHM, being $D_p$ the dose per pulse, while $\dot{D}_{av}$ is $D_{p}\cdot f$. Based on $E$, the particles stopping power ($S$/$\rho$) is interpolated from the NIST Pstar database\cite{berger1998stopping}, and the number of primary particles ($N$) is obtained from Eq. \eqref{eq:5}.
\begin{equation}
D = \frac{N}{L^2} \cdot \frac{S}{\rho}
\label{eq:5}
\end{equation}
After completing these initial calculations, the simulation kicks off with the physical stage for all $N$. These are created within a square of side length $L$ (VOI characteristic side length) on the XY plane inside a 1 km$^3$ mother volume, so all the secondary electrons are contained. The VOI, which is designated for the physico-chemical  and non-homogeneous chemical stages, is situated at the origin of the mother volume. To account for the build-up within the VOI, the primaries are created at a depth $-z_0$ larger than the range of a delta electron with an energy equal to the maximum energy transfer ($Q$) between the primary and the electron. Similarly, to account for the backscattering, they are terminated after they  loose kinetic energy exceeding $Q$, well beyond the VOI. The energy depositions from the primary particles and the secondary ones, generated inside yet leaving the VOI, are recorded. The collection of all the VOI containing both primary and secondary energy deposits are treated as independent radiation events throughout the beam-on time. These are randomly sampled in time according to the dose rates and the pulse beam structure. This setup emulates the impact of scattered radiation coming from outside the VOI. Using this methodology, the unrestricted LET (LET$_\infty$), can be calculated as $D \cdot \rho \cdot L^2 / N$, matching the $S/\rho$ of the primaries. This approach was similarly used in RITRACKS v3.07 with a PBC, which was used to account for delta electrons from neighboring volumes \cite{brogan2015benchmarking}.

Following, the physico-chemical  stage for all the radiation events is performed individually and the positions of the first chemical products are recorded. Both the physical stage and the physico-chemical stage are considered instantaneous throughout the beam-on time (non-homogeneous stage), during which, gMicroMC's SBS along with the newly incorporated PBCs (Section \ref{pbc}) are active (SBS + PBCs). At this stage, radiation is coming in, and the initial chemical products from each radiation event are incorporated according to the events-time distribution. Subsequently, during the remainder of the beam-off time, or homogeneous stage, the NumODEs are performed until the next pulse arrives (Sec. \ref{NumODEs}). The alternation of SBS + PBCs and NumODEs is repeated for each pulse. After the last pulse, the NumODEs are run for 10 s, so a steady-state in the concentration of species is reached. Finally, the post-irradiation G-values are recorded.

%%%%%%%%%%%%%%%%%%%%%%%%%%%%%%%%%%%%%%%%%%%%%%%%%%%%%%%%%%%%%%
%%%%%%%%%%%%%%%%%%%%%%%%% 2D Lattice %%%%%%%%%%%%%%%%%%%%%%%%%
%%%%%%%%%%%%%%%%%%%%%%%%%%%%%%%%%%%%%%%%%%%%%%%%%%%%%%%%%%%%%%
\vspace{0.5cm}
\begin{figure}[H]
   \begin{center}
   \includegraphics[height=6cm]{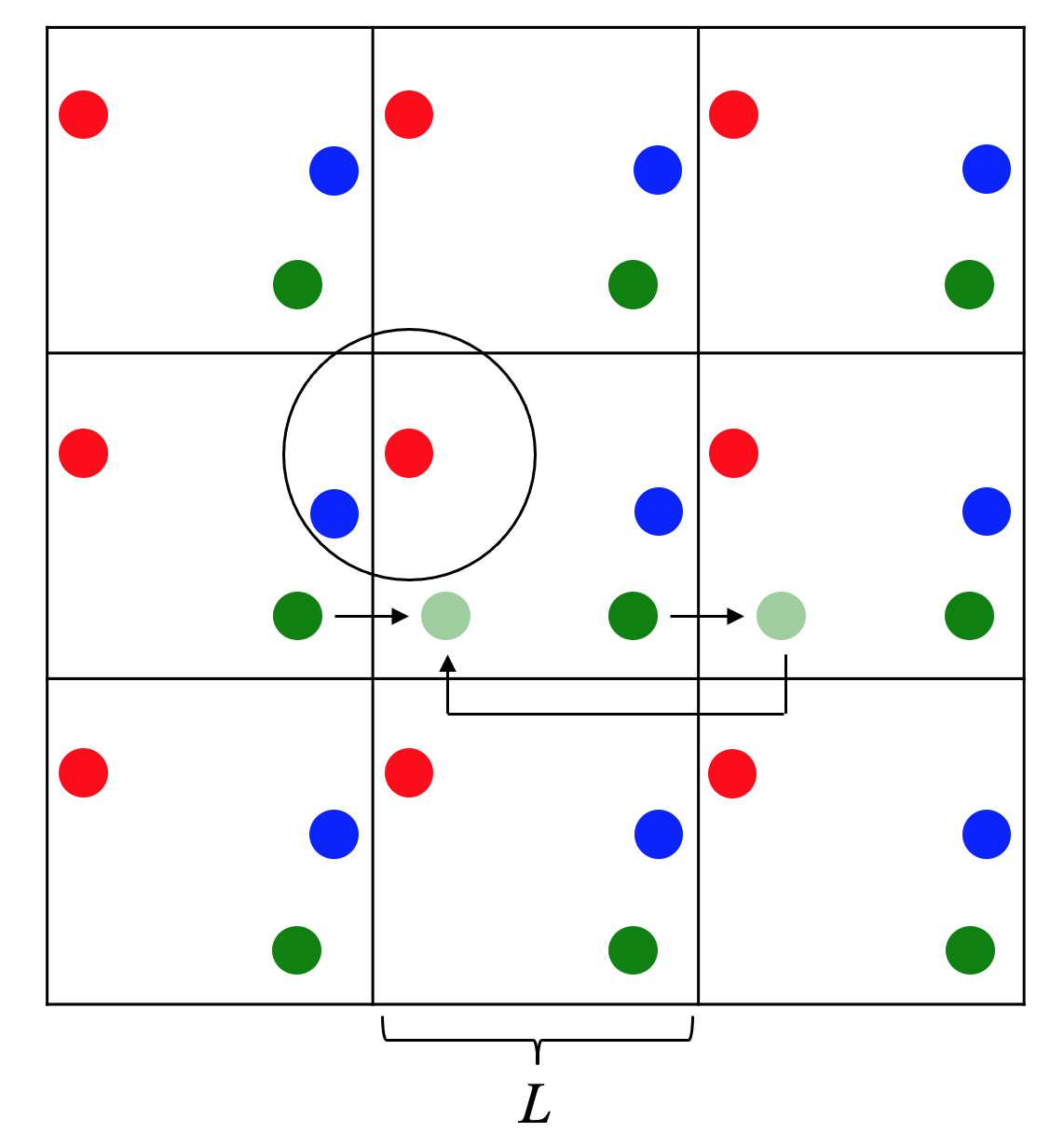}
   \caption
   {Two-dimensional periodic lattice displaying three hypothetical chemicals (red, green, blue) per voxel. The green chemical diffuses out of the voxel and re-enter from the opposite side (arrows) due to periodic boundary condition 1. The red and blue chemicals, from different voxels, are subjected to periodic boundary condition 2 to react as they fall within the reaction radius (circle).   
   }
   \label{PBClattice} 
   \end{center}
\end{figure}

%%%%%%%%%%%%%%%%%%%%%%%%%%%%%%%%%%%%%%%%%%%%%%%%%%%%%%%%%%%%%%
%%%%%%%%%%%%%%%%%%%%%%%%%%% PBCs %%%%%%%%%%%%%%%%%%%%%%%%%%%%%
%%%%%%%%%%%%%%%%%%%%%%%%%%%%%%%%%%%%%%%%%%%%%%%%%%%%%%%%%%%%%%

\subsection{Theory of periodic boundary conditions}
\label{pbc}
This standard technique from Molecular Dynamics (MD)\cite{frenkel2000molecular} was included in gMicroMC’s SBS algorithm. PBCs are responsible of ensuring that the VOI accurately reflects the bulk's properties. They are instrumental in bridging the divide between the traditional microscopic calculations of independent chemical tracks and the simulation of the macroscopic chemical yield. According to available literature, the only MCTS algorithm to ever include PBCs for the non-homogeneous chemical stage was RITRACKS \cite{plante2017considerations} to compare the independent reaction time and SBS methods. The theoretical approach of PBCs envisions the bulk system as an infinite lattice of periodic identical cubic boxes, or voxels, each with a side length $L$ (Fig. \ref{PBClattice}). In MCTS modeling, a key input is the VOI, defined by $L$, which essentially represents a single voxel within the infinite lattice. The PBCs regulates the influx and efflux of chemicals at the boundaries of the VOI. Additionally, they should guarantee the reactions between reactants in adjacent voxels when they lie within the reaction radius ($\mathcal{R}$). To ensure an accurate handling of surface effects and that the representativeness of the bulk is guaranteed, two PBCs were implemented.

\noindent – \textbf{Periodic boundary condition 1}: Within the infinite grid, if a chemical exits a voxel, its image reenters on the opposite side, like the green chemical in Fig. \ref{PBClattice}. After every diffusion jump, this PBC assesses the chemicals positions ($x$) to check whether any of them exited the boundaries. If a chemical exits the VOI, Equation \eqref{eq:1} recalculates $x$, so it reenters on the opposite side. \begin{equation}
x = \begin{cases} 
x - L & \text{if } x \geq \frac{L}{2}, \\
x + L & \text{if } x < -\frac{L}{2}.
\end{cases}
\label{eq:1}
\end{equation}

\noindent – \textbf{Periodic boundary condition 2}: In Fig. \ref{PBClattice}, consider the red reactant and the image of the blue reactant, which is in the nearby box. They could potentially react as they fall within $\mathcal{R}$ (circle), but the image of the blue reactant is out of the scope of the simulation. To facilitate these reactions, this PBC substitutes the interaction between the red reactant and the image of the blue reactant with the actual position of the blue reactant, which is situated on the opposing side of the voxel and within the simulated region. This adjustment is made by recalculating the distance ($dx$) between the red and blue reactants (Eq. \eqref{eq:2}). To prevent reactions between a reactant and its own image, the minimum image convention stipulates that the ratio between $\mathcal{R}$ and $L$ should satisfy $2\mathcal{R} < L$. \begin{equation}
dx = \begin{cases} 
dx - L & \text{if } dx \geq \frac{L}{2}, \\
dx + L & \text{if } dx < -\frac{L}{2}.
\end{cases}
\label{eq:2}
\end{equation}

%%%%%%%%%%%%%%%%% Setup PBCs verification %%%%%%%%%%%%%%%%%%%%%

% \subsubsection{Setup for verification of periodic boundary conditions}
When using PBCs, there are several sources of potential systematic errors, known as finite-size effects \cite{frenkel2000molecular}. These include an insufficient number of chemicals, volume related factors, and potential bugs in the software code. To demonstrate the correct implementation of the PBCs, it is required to properly manage any finite-size effects to ultimately show that the VOI is representative of an infinite volume\cite{allen2017computer}. Volume-scaling trials are a standard practice within MD to verify this. The species concentration ($C$) time evolution was examined for an homogeneous mixture across VOI characterized by $L$ with values of 0.2, 0.5, and 1.0 $\muu$m. The initial $C$ was proportional to the G$_0$ produced by a single 70 MeV proton pulse characterized by FWHM = 1 $\muu$s and $\dot{D}_p$ = 10$^{6}$ Gy/s. The initial $C$ and the $L$ sizes were strategically made to ensure that a steady state in $C$ was achieved. Larger VOI with same initial $C$ would not be computationally viable because it would require tracing a vast number of chemicals. Whereas if the initial $C$ was decreased by enlarging the VOI, the time to reach the steady state would be too large. The volume independence of $C$ was examined by comparison against the analytical solution from Kinetiscope\cite{wiegel2015stochastic}, which uses the Gillespie algorithm\cite{gillespie1977exact} to calculate $C$. The ratio between the two methods was evaluated.

%%%%%%%%%%%%%%%%%%%%%%%%%%%%%%%%%%%%%%%%%%%%%%%%%%%%%%%%%%%%%%
%%%%%%%%%%%%%%%%%%%%%%%%%%% NumODEs %%%%%%%%%%%%%%%%%%%%%%%%%%
%%%%%%%%%%%%%%%%%%%%%%%%%%%%%%%%%%%%%%%%%%%%%%%%%%%%%%%%%%%%%%

\subsection{Homogeneous stage modeling}
\label{NumODEs}
To simulate the homogeneous stage, alternatively the beam-off time, the chemicals need to be monitored for periods of time that extend from the $\muu$s scale to seconds and minutes. The SBS is computationally inefficient for tracing the chemicals over these very large periods. To tackle this issue, the NumODEs were incorporated into the gMicroMC, providing a speed up factor several orders of magnitude faster than the SBS.  The empirical differential rate equations were solved with a numerical discretization method following the method from Labarbe et al\cite{labarbe2020physicochemical}. 
\begin{equation}
\frac{d[e_{\text{aq}}^-]}{dt} = -2\cdot R_{1} - R_{2} - R_{3} - R_{4} - R_{5}
\label{eq:3}
\end{equation}
\begin{equation}
R_{1}=k_{1}\cdot[e_{\text{aq}}^-]\cdot[e_{\text{aq}}^-]
\label{eq:4}
\end{equation}%

For example, the consumption rate of e$_{\text{aq}}^-$ is given to the ordinary differential equation shown in Eq. \ref{eq:3}, where each term corresponds to the reaction rate from each reaction involving e$_{\text{aq}}^-$. For instance, Eq. \ref{eq:4} represents the reaction rate corresponding to R1 (Tbl. \ref{reactionrates}). All of these ordinary differential equations were numerically integrated. 

To benchmark the goodness of the NumODEs, a cross-validation with Kinetiscope was conducted following the setup of the PBCs validation. A mixture proportional to the G$_0$ produced by a single 70 MeV proton pulse characterized by FWHM = 1 $\muu$s and $\dot{D}_p$ = 10$^{6}$ Gy/s was homogeneously mixed. The $C$ ratio between Kinetiscope and  NumODEs for every specie was examined.

%%%%%%%%%%%%%%%%%%%%%%%%%%%%%%%%%%%%%%%%%%%%%%%%%%%%%%%%%%%%%%
%%%%%%%%%%%%%%%%%%%%% gMicroMC Updates %%%%%%%%%%%%%%%%%%%%%%%
%%%%%%%%%%%%%%%%%%%%%%%%%%%%%%%%%%%%%%%%%%%%%%%%%%%%%%%%%%%%%%

\subsection{Physico-chemistry and chemistry updates}
\label{methodsupdates}
Based on Geant4-DNA\cite{shin2021geant4}, new dissociation schemes and branching ratios were added to gMicroMC's physico-chemical module (Table \ref{branchingratios} in Supplement). For the chemistry modeling, the radiolysis literature was revisited, seeking experimental measurements of reaction rates ($k$) and of the species diffusion coefficients ($\mathcal{D}$). The chemistry list was updated based on Elliot et al.\cite{elliot1994rate}. The reaction list with their respective $k$ are shown in Tbl. (\ref{reactionrates}), while the $\mathcal{D}$ are listed in Tbl. \ref{DiffusionContinued} (Supplement). For self-radical recombination reactions (R1, R6, R8), $\mathcal{R}$ was calculated with an input reaction rate of 2$k$. According to the Smoluchowski formalism \cite{rice1985diffusion}, $\mathcal{R}$ and $k$ are linked by equaling Fick's first law with the empirical differential rate equation. This equation includes a factor of 2 for self-interacting reactants, which should be considered when calculating $\mathcal{R}$. A detailed analysis can be found in the Supplement.
\vspace{0.75cm}
\begin{table}[H]
\small
\centering
\begin{tabular}{c|c|c|}
\cline{2-3}
                         & \textbf{Reaction}  \rule{0pt}{3ex}                                                     & \textbf{$k$ (10$^{10}\cdot$l$\cdot$mol$^{-1}\cdot$s$^{-1}$)} \rule{0pt}{3ex} \\ \hline
\multicolumn{1}{|c|}{\textbf{R1}} & e$_{\text{aq}}^-$ + e$_{\text{aq}}^-$ \ce{->} H$_2$ + 2OH$^{-}$ & 2$k$ = 1.3                                       \\ \hline
\multicolumn{1}{|c|}{\textbf{R2}} & e$_{\text{aq}}^-$ + $^{\cdot}$OH \ce{->} OH$^{-}$               & 3.0                                               \\ \hline
\multicolumn{1}{|c|}{\textbf{R3}} & e$_{\text{aq}}^-$ + H$^{\cdot}$ \ce{->} H$_2$ + OH$^{-}$        & 2.6                                               \\ \hline
\multicolumn{1}{|c|}{\textbf{R4}} & e$_{\text{aq}}^-$ + H$^{+}$ \ce{->} H$^{\cdot}$                 & 2.3                                               \\ \hline
\multicolumn{1}{|c|}{\textbf{R5}} & e$_{\text{aq}}^-$ + H$_2$O$_2$ \ce{->} $^{\cdot}$OH + OH$^{-}$  & 1.4                                               \\ \hline
\multicolumn{1}{|c|}{\textbf{R6}} & $^{\cdot}$OH + $^{\cdot}$OH \ce{->} H$_2$O$_2$                  & 2$k$ = 0.95                                       \\ \hline
\multicolumn{1}{|c|}{\textbf{R7}} & $^{\cdot}$OH + H$^{\cdot}$ \ce{->} H$_2$O                       & 1.5                                               \\ \hline
\multicolumn{1}{|c|}{\textbf{R8}} & H$^{\cdot}$ + H$^{\cdot}$ \ce{->} H$_2$                         & 2$k$ = 1.1                                       \\ \hline
\multicolumn{1}{|c|}{\textbf{R9}} & H$^{+}$ + OH$^{-}$ \ce{<=>} H$_2$O                               & 11.                                                \\ \hline
\end{tabular}
\caption{Updated gMicroMC chemical reactions with their $k$ collected from Elliot et al.\cite{elliot1994rate}}
\label{reactionrates}
\end{table}
\vspace{0.5cm}
%%%%%%%%%%%%%%%%% Setup updates verification %%%%%%%%%%%%%%%%

To determine the chemical yields of independent tracks shortly after their creation, specifically at times ($t$) below 1 $\muu$s, pulse radiolysis experimentalists employ pulses with durations in the same order of magnitude as $t$. For example, pulses with FWHM = 1.4 $\muu$s were used to characterize the H$_2$ and H$_2$O$_2$ yields at the $\muu$s scale \cite{anderson1962radiation}. Similarly, ns pulses were utilized to measure  e$_{\text{aq}}^-$\cite{buxton1972nanosecond}. Other more precise investigations measured $^{\cdot}$OH at 10 ps and 100 ps\cite{el2011time, jay2000new}. MCTS models are validated by simulating independent chemical tracks up to 1 $\muu$s, and comparing the G$_0$ time evolution against pulse radiolysis experimental measurements and scavenging data. With the purpose of validating the new physico-chemistry and chemistry lists, single pulse radiolysis was simulated in a VOI with $L$ = 5 $\muu$m. Firstly, single 70 MeV protons pulses with FWHM values of 10 ps, 1 ns and 1 $\muu$s were simulated, each with respective $\dot{D}_p$ of 10$^{11}$, 10$^{9}$, and 10$^{6}$ Gy/s. Paralelly, single pulses with proton energies of 1, 3, 10, 25, 70, and 224 MeV were simulated, which were characterized by FWHM = 1 $\muu$s and $\dot{D}_p$ = 10$^6$ Gy/s. The G-values right after radiation stopped were compared against experimental data for both setups.

%%%%%%%%%%%%%%%%%%%%%%%%%%%%%%%%%%%%%%%%%%%%%%%%%%%%%%%%%%%%%%
%%%%%%%%%%%%%%%%%%%%%%%%% Set-up %%%%%%%%%%%%%%%%%%%%%%%%%%%%%
%%%%%%%%%%%%%%%%%%%%%%%%%%%%%%%%%%%%%%%%%%%%%%%%%%%%%%%%%%%%%%

\subsection{Multiple pulse radiolysis setup at SDR and UHDR}
\label{dosimetric_setup}
To study the dose rate dependencies on radical-radical reactions, the used pulse beam structure replicated that used in experiments where the Flash effect was observed\cite{vozenin2019biological}. These are characterized by a FWHM in the $\muu$s scale, and a $\dot{D}_p$ around 10$^5$ – 10$^6$ Gy/s. The $\dot{D}_{av}$ is adjusted by modifying $f$. This set-up is complementary with the standard pulse radiolysis experiments, which employ pulses with widths in the ns – $\muu$s order and deliver about 1 Gy per pulse\cite{wardman2020radiotherapy}. While such structures are typical for electron beams, protons were used in this simulation study. The G-values are primarily dependent LET rather than the radiation type, therefore electrons and protons with similar LET are expected to produce equivalent G-values. In fact, below 1 keV/$\muu$m the G$_0$ remains relatively consistent\cite{ramos2018monte}.

In this work, protons with 70 MeV of kinetic energy, which have an LET = 0.96 keV/$\muu$m, were primarily used since they present several computational advantages. First, they travel in straighter paths than electrons, which follow more erratic trajectories and could exit micrometric VOI. Secondly, 70 MeV protons have an LET four or five greater than clinical electrons, which have an LET  around 0.2 keV/$\muu$m, hence between four and five less protons are required to deposit the same dose (Eq. \ref{eq:1}). Using 70 MeV protons produces a similar chemical yield as clinical electrons but with a reduction of the computational load and an efficiency increase. Specifically, the simulations consisted of pulses characterized by FWHM = 1 $\muu$s. The total dose amounted up to 20 Gy. The VOI was defined by $L$ = 2 $\muu$m. The $\dot{D}_{p}$ values were 2$\cdot$10$^5$ and 10$^6$ Gy/s, while the $\dot{D}_{av}$ was adjusted by changing the pulse $f$, ranging from 0.01 to 100000 Gy/s.

%%%%%%%%%%%%%%%%%%%%%%%%%%%%%%%%%%%%%%%%%%%%%%%%%%%%%%%%%%%%%%
%%%%%%%%%%%%%%%%%%%%%%%%%%%%%%%%%%%%%%%%%%%%%%%%%%%%%%%%%%%%%%
%%%%%%%%%%%%%%%%%%%%%%%%%%%%%%%%%%%%%%%%%%%%%%%%%%%%%%%%%%%%%%
%%%%%%%%%%%%%%%%%%%%%%%%% Results  %%%%%%%%%%%%%%%%%%%%%%%%%%%
%%%%%%%%%%%%%%%%%%%%%%%%%%%%%%%%%%%%%%%%%%%%%%%%%%%%%%%%%%%%%%
%%%%%%%%%%%%%%%%%%%%%%%%%%%%%%%%%%%%%%%%%%%%%%%%%%%%%%%%%%%%%%
%%%%%%%%%%%%%%%%%%%%%%%%%%%%%%%%%%%%%%%%%%%%%%%%%%%%%%%%%%%%%%

\section{Results}

%%%%%%%%%%%%%%%%%%%%%%%%%%%%%%%%%%%%%%%%%%%%%%%%%%%%%%%%%%%%%%
%%%%%%%%%%%%%%%%%%%%%%%% PBCs verification %%%%%%%%%%%%%%%%%%%
%%%%%%%%%%%%%%%%%%%%%%%%%%%%%%%%%%%%%%%%%%%%%%%%%%%%%%%%%%%%%%

\subsection{Verification of periodic boundary conditions and numerical ordinary differential equations}
\begin{figure}[H]
   \begin{center}
   \includegraphics[width=16.5cm]{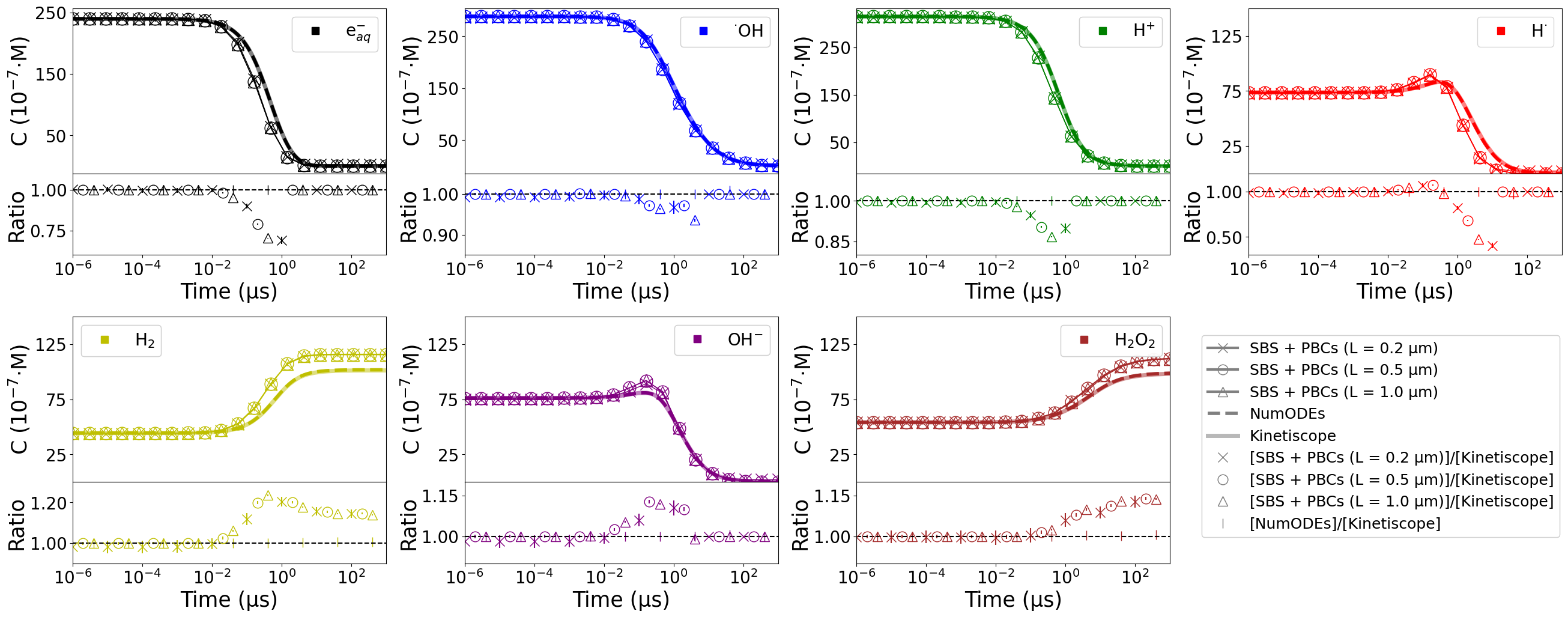}
   \caption{(\textbf{Upper subplots}) Concentration time evolution calculated using the SBS + PBCs, NumODEs, and Kinetiscope for a homogeneous mixture proportional to the G$_0$ of single 70 MeV proton pulse, characterized by FWHM = 1 $\mu$s and $\dot{D}_{p}$ = 10$^6$ Gy/s. The SBS + PBCs calculations are within VOI characterized by \textit{L} equal to 0.2, 0.5, and 1.0 $\mu$m. \textbf{(Lower subplots)} Ratio of concentration calculated by the SBS + PBCs within each VOI with Kinetiscope ([SBS + PBCs]/[Kinetiscope]), and NumODEs with Kinetiscope ([NumODEs]/[Kinetiscope]).
    }
    \label{PBC}
    \end{center}   
\end{figure}

Regarding the PBCs validation, there were not significant differences among the three VOI tested with the SBS + PBCs: all of them predicted the same $C$. This indicates that no finite-size were encountered, and that the PBCs do not bias the SBS (Fig. \ref{PBC}). Although both the SBS + PBCs and Kinetiscope show the same tendency for the depletion and creation of chemicals, there were significant distinctions. When the steady state was achieved at 1 ms, the maximum concentration differences were about 20\% and 12\% for H$_2$ and H$_2$O$_2$, respectively. These disparities arised because R1 predominated in the SBS + PBCs compared to Kinetiscope. When this reaction dominates, the production of H$_2$ increases. Consequently, R2 slows down, leading to a higher production of H$_2$O$_2$ via R6.

%%%%%%%%%%%%%%%%%%%%%%%%%%%%%%%%%%%%%%%%%%%%%%%%%%%%%%%%%%%%%%
%%%%%%%%%%%%%%%%%%%%%%% PBCs and NumODEs %%%%%%%%%%%%%%%%%%%%%
%%%%%%%%%%%%%%%%%%%%%%%%%%%%%%%%%%%%%%%%%%%%%%%%%%%%%%%%%%%%%%
\vspace{0.5cm}

The differences between the SBS + PBCs and Kinetiscope were attributed to the fundamental differences between the two methods: the first is particle-based, while the other uses the Gillespie algorithm, which is a stochastic simulation algorithm. The Gillespie algorithm randomly selects which reaction will occur next and calculates the time until this reaction happens, based on reaction probabilities derived from the $k$ list. On the other hand, the SBS algorithm correlates $\mathcal{R}$, $k$ and $\mathcal{D}$; a reaction takes places when two reactants fall within $\mathcal{R}$. Additionally, the SBS models are subject to uncertainties due to unrefined parameter values as the $\mathcal{D}$\cite{lai2021modeling}. Beyond merely affirming the PBCs, this validation provides a more comprehensive validation of the SBS algorithm than the conventional verification. Traditionally, the SBS algorithms are benchmarked against G$_0$ experimental measurements, a complex approach since physics, physico-chemistry, and chemistry are involved. This validation is significantly more straightforward, as it solely focuses on chemistry. Only because the PBCs ensures the representativeness of the bulk, this comparison with Kinetiscope is possible, which similarly models the macroscopic properties of homogeneous mixtures.

Both the NumODEs and Kinetiscope predicted the same $C$, and the ratio for all the species at any given time was below 1\% (Fig. \ref{PBC}). The NumODEs serve as an optimal tool for modeling the homogeneous stage since it provides a speed-up factor several orders of magnitude above the SBS. For instance, the previous evaluation took approximately 1 day using the SBS + PBCs ($L$ = 1.0 $\muu$m), whereas the NumODEs and Kinetiscope completed the same task in less than a second, providing a speed-up factor of $\sim$86400.

The multiple pulse extension of gMicroMC operates by alternating the SBS + PBCs for the non-homogeneous stage, and the NumODEs for the homogeneous stage. We acknowledge that combining a particle-based method with a numerical one might introduce limitations, as the results in this section indicate significant distinctions between the two. However, it is a step forward in the simulation of realistic irradiations and future works will isolate the differences between these approaches.

%%%%%%%%%%%%%%%%%%%%%%%%%%%%%%%%%%%%%%%%%%%%%%%%%%%%%%%%%%%%%%
%%%%%%%%%%%%%%%%%%%%%% Single pulse %%%%%%%%%%%%%%%%%%%%%%%%%%
%%%%%%%%%%%%%%%%%%%%%%%%%%%%%%%%%%%%%%%%%%%%%%%%%%%%%%%%%%%%%%

\subsection{Verification of physico-chemistry and chemistry updates}
\begin{figure}[H]
   \begin{center}
   \includegraphics[width=15cm]{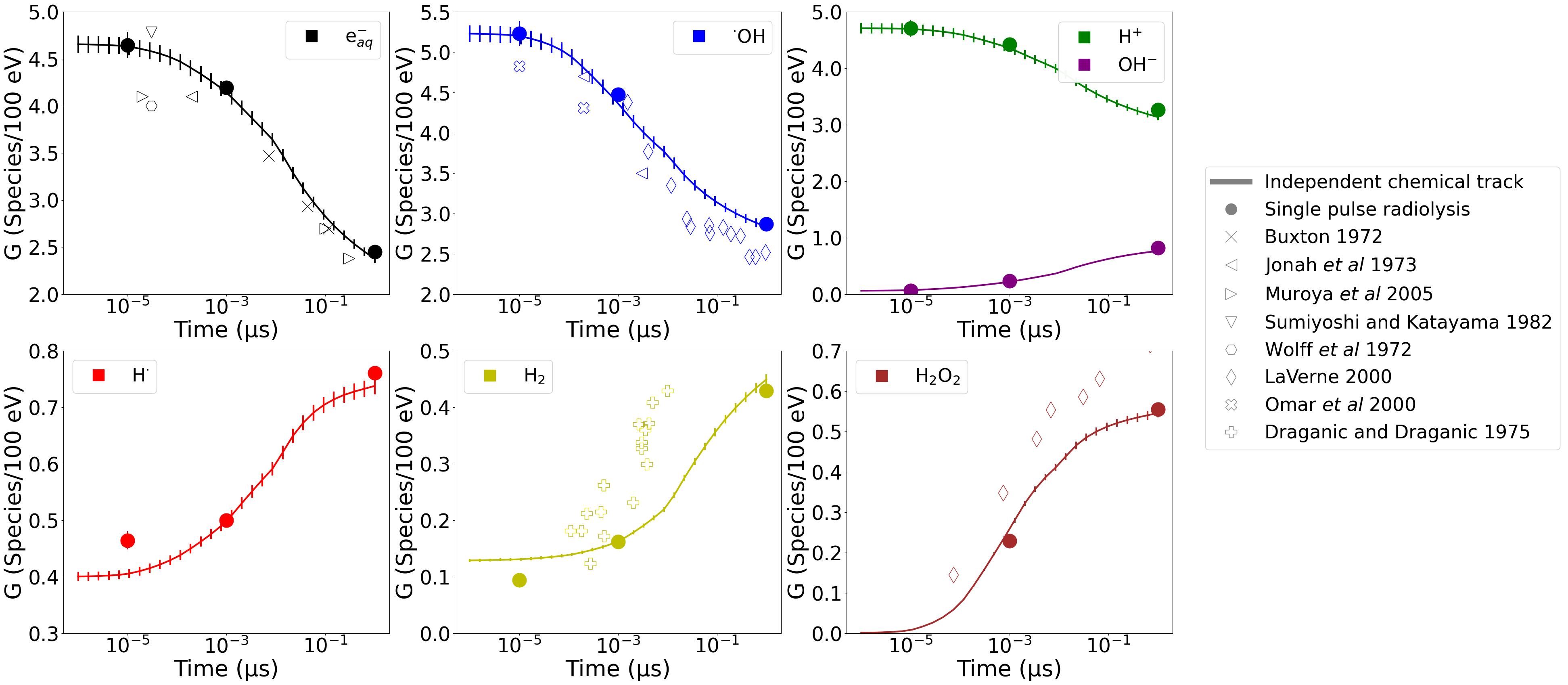}
   \caption
   {G$_0$ of chemical species up to 1 $\muu$s for 70 MeV protons. It shows the simulation of an independent chemical track, and the yield after radiation stops for single pulse radiolysis simulations with FWHM values of 10 ps, 1 ns, and 1 $\muu$s, with respective $\dot{D}_p$ of 10$^{11}$, 10$^{9}$, and 10$^{6}$ Gy/s. The symbols are experimental measurements\cite{buxton1972nanosecond, jonah1973yields,muroya2005re,sumiyoshi1982yield,wolff1973picosecond,laverne2000oh,el2011time, draganic1975formation}.
    }  
    \label{fig:SinglePuseRadiolysisFWHM} 
    \end{center}
\end{figure}

Fig. \ref{fig:SinglePuseRadiolysisFWHM} compares the G$_0$ for the three 70 MeV proton pulses simulated (full circles), the simulation of an independent 70 MeV proton track (solid line), and experimental data. The G$_0$ measured after the single pulse simulations follows the tendency of the independent chemical track, thus demonstrating that the model accurately reflects the macroscopic nature of pulse radiolysis by averaging the multiple chemical tracks that coexist within each pulse. 

\begin{figure}[H]
   \begin{center}
   \includegraphics[width=14cm]{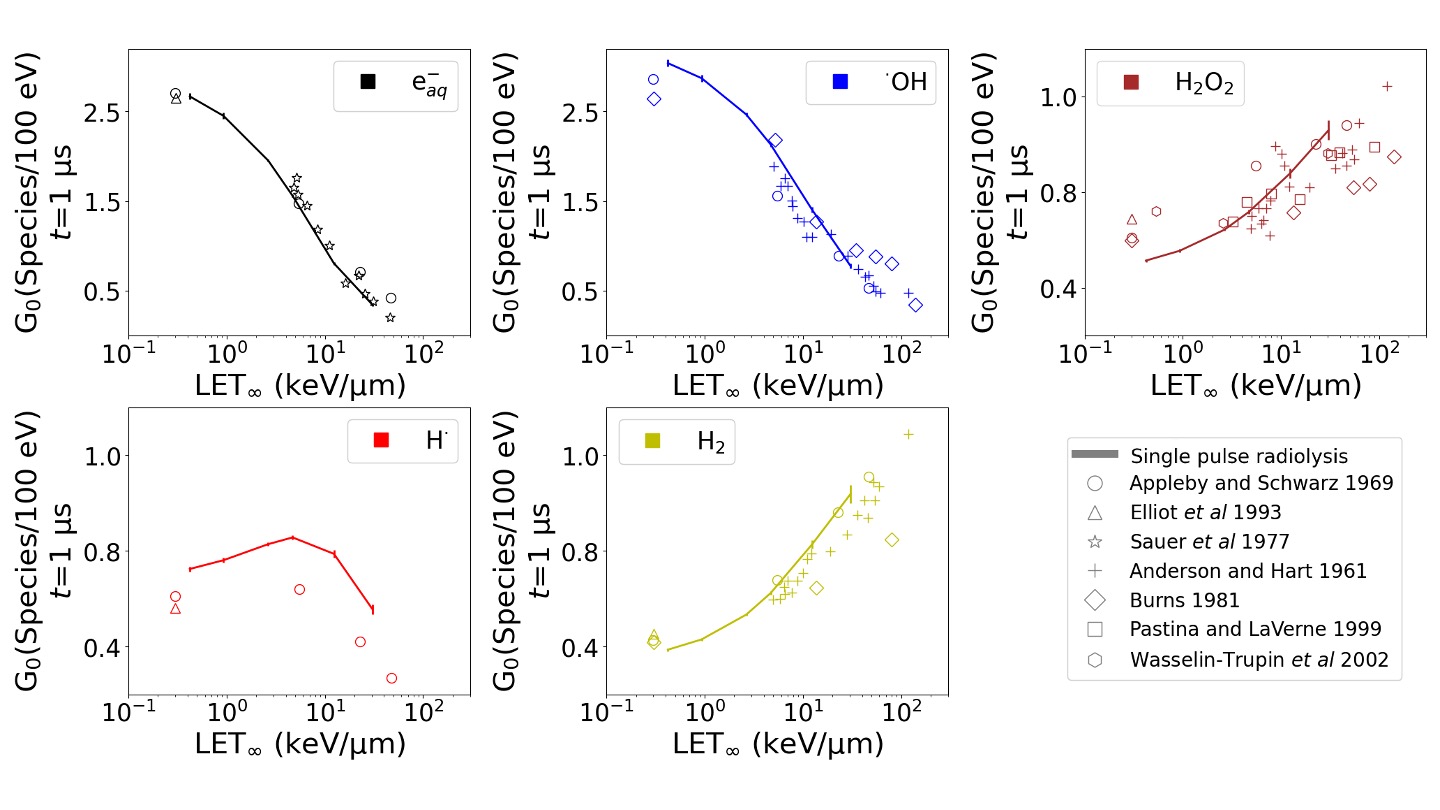}
   \caption{G$_0$ of different species by a single pulse with FWHM = 1 $\muu$s and $\dot{D}_{p}$ = 10$^6$ Gy/s of protons with increasing LET$_\infty$ (straight lines with error bars). The symbols are experimental measurements\cite{appleby1969radical, johnaelliot1993temperature, sauer1977let, anderson1961molecular, burns1981effect, pastina1999hydrogen, wasselin2002hydrogen}.}  %note label inside caption
    \label{LET} 
    \end{center}
\end{figure}

The G$_0$ as a function of LET$_\infty$ is presented in Fig. \ref{LET} among experimental measurements. At high LET ($\geq$ 1 keV/$\muu$m), energy depositions are densely concentrated along the tracks, which intensifies the accumulation of the first chemicals species, thus enhancing chemical kinetics. As a result, radicals deplete rapidly while the production of molecules increases. Additionally, the LET$_\infty$ within the VOI matched the protons $S$/$\rho$ (Fig. \ref{Nist} in Supplement).  The energy deposits by delta electrons escaping the VOI ($L$ = 5 $\muu$m), which were reflected during the beam-on time, were up to 20\% of the total energy lost by the primary protons. 

\vspace{0.5cm}

\vspace{-0.5cm}

In this section, the cross-validation between MCTS and pulse radiolysis experiments was enhanced by accurately reproducing the macroscopic characteristics of pulse radiolysis experiments. The physico-chemical and chemical updates are validated since the pulse radiolysis experimental tendency was replicated.

%%%%%%%%%%%%%%%%%%%%%%%%%%%%%%%%%%%%%%%%%%%%%%%%%%%%%%%%%%%%%%
%%%%%%%%%%%%%%%%%%%%%% Multiple pulse %%%%%%%%%%%%%%%%%%%%%%%%
%%%%%%%%%%%%%%%%%%%%%%%%%%%%%%%%%%%%%%%%%%%%%%%%%%%%%%%%%%%%%%

\subsection{Multiple pulse radiolysis at SDR and UHDR}
\begin{figure}[H]
   \begin{center}
   \includegraphics[width=16.5cm]{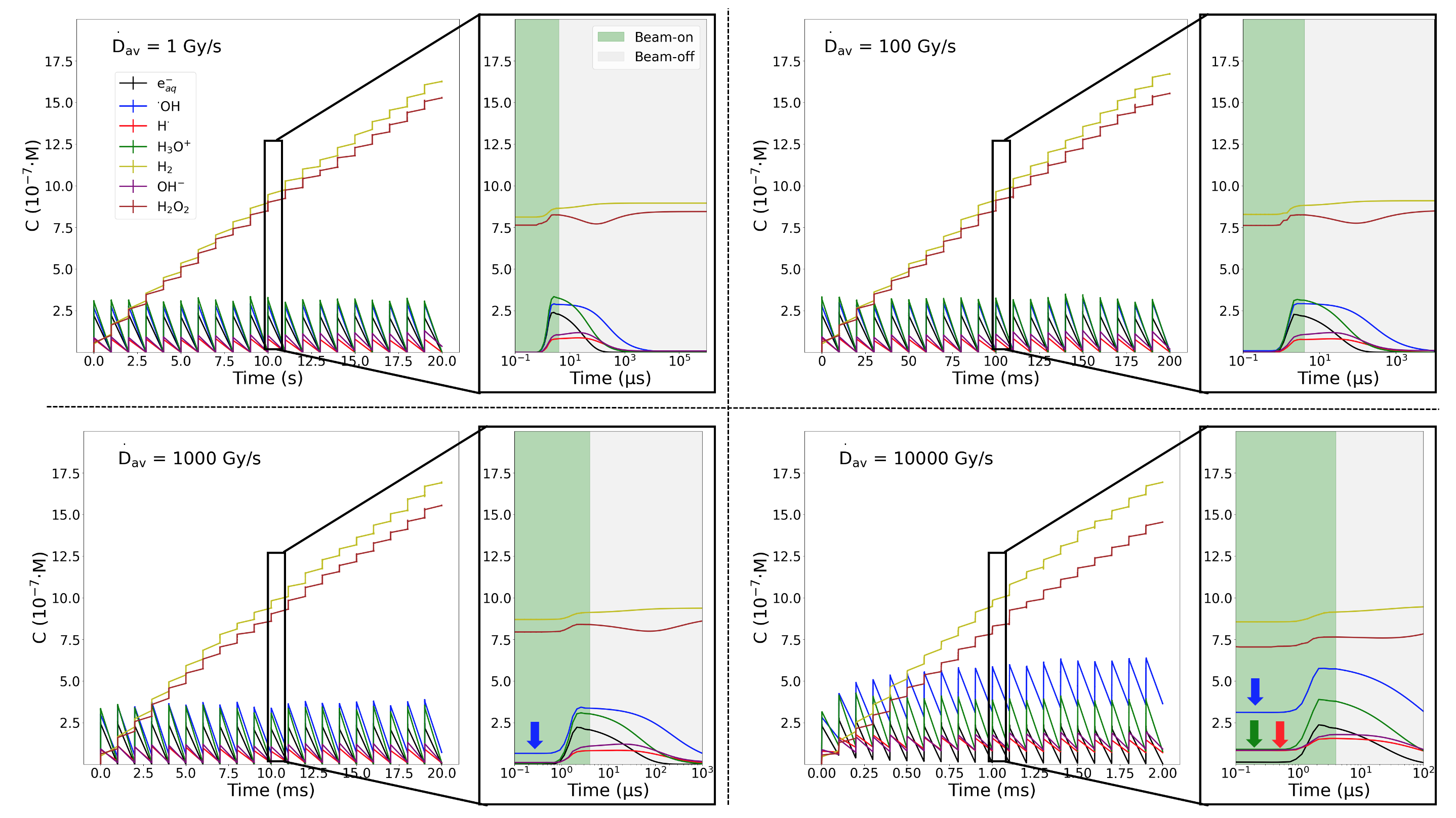}
   \caption
   {Time dependence of species concentration during a twenty 70 MeV proton irradiation characterized by FWHM = 1 $\muu$s and $\dot{D}_{p}$ = 10$^6$ Gy/s. Plots represent specific irradiations at varying $\dot{D}_{av}$ of 1, 100, 1000 and 10000 Gy/s. A zoomed-in view in logarithmic scale of the tenth pulse shows the beam-on time (non-homogeneous stage), the beam-off time (homogeneous stage) and reactive species build-up at the inception of the pulses marked as arrows.   
    }  %note label inside caption
    \label{Concentration-Time} 
    \end{center}
\end{figure}

 Fig. \ref{Concentration-Time} presents the species $C$ time evolution for irradiations characterized by FWHM = 1 $\muu$s and $\dot{D}_{p}$ = 10$^6$ Gy/s. Each case shows $\dot{D}_{av}$ of 1, 100, 1000 and 10000 Gy/s. The spikes in $C$ correspond to every radiation pulse. As the dose escalates the H$_2$ and H$_2$O$_2$ molecules accumulate, while the rest of the radicals are consumed. Between 100 Gy/s and 1000 Gy/s the steady state in the species $C$ stops being achieved within the pulses and, multiple species start accumulating throughout the irradiation. This accumulation is the reactive species build-up, and it is marked with arrows in the zoomed-in sections at the plots, which detail the creation of chemicals during the non-homogeneous phase and their subsequent consumption during the homogeneous stage at the tenth pulse.
 
 %%%%%%%%%%%%%%%%%%%%%%%%%%%%%%%%%%%%%%%%%%%%%%%%%%%%%%%%%%%%%%
%%%%%%%%%%%%%%%%%%%%%% Time-Concentration SDR %%%%%%%%%%%%%%%%
%%%%%%%%%%%%%%%%%%%%%%%%%%%%%%%%%%%%%%%%%%%%%%%%%%%%%%%%%%%%%%
\vspace{0.5cm}
 
 A discernible $\dot{D}_{av}$ threshold is evident, marking the critical value at which the chemical species shift from reaching a steady state within the pulses to  a non-steady state regime. Below the threshold (SDR), the steady state is consistently achieved; above it (UHDR), the steady state is not attained and reactive species accumulate. While at SDR the only species that accumulate at the end of the pulses are H$_2$ and H$_2$O$_2$, under UHDR conditions, the shorter intervals between pulses prevent the achievement of the steady state, leading to the build-up of multiple species types, including $^{\cdot}$OH radicals. As the $\dot{D}_{av}$ increases, the time between pulses decreases, making richer the reactive species build-up: the radicals have less time to react. This enhanced reactive species build-up is the driver of the radical-radical reactions promotion at UHDR. Specifically, for $\dot{D}_{p}$ = 2·10$^5$ Gy/s, the $\dot{D}_{av}$ threshold was in the order of 10 Gy/s, while for $\dot{D}_{p}$ = 10$^6$ Gy/s it was identified one factor above, about 100 Gy/s.

%%%%%%%%%%%%%%%%%%%%%%%%%%%%%%%%%%%%%%%%%%%%%%%%%%%%%%%%%%%%%%
%%%%%%%%%%%%%%%%%%%% Concentration-Dose %%%%%%%%%%%%%%%%%%%%%%
%%%%%%%%%%%%%%%%%%%%%%%%%%%%%%%%%%%%%%%%%%%%%%%%%%%%%%%%%%%%%%
\vspace{0.5cm}
\begin{figure}[H]
   \begin{center}
   \includegraphics[height=6cm]{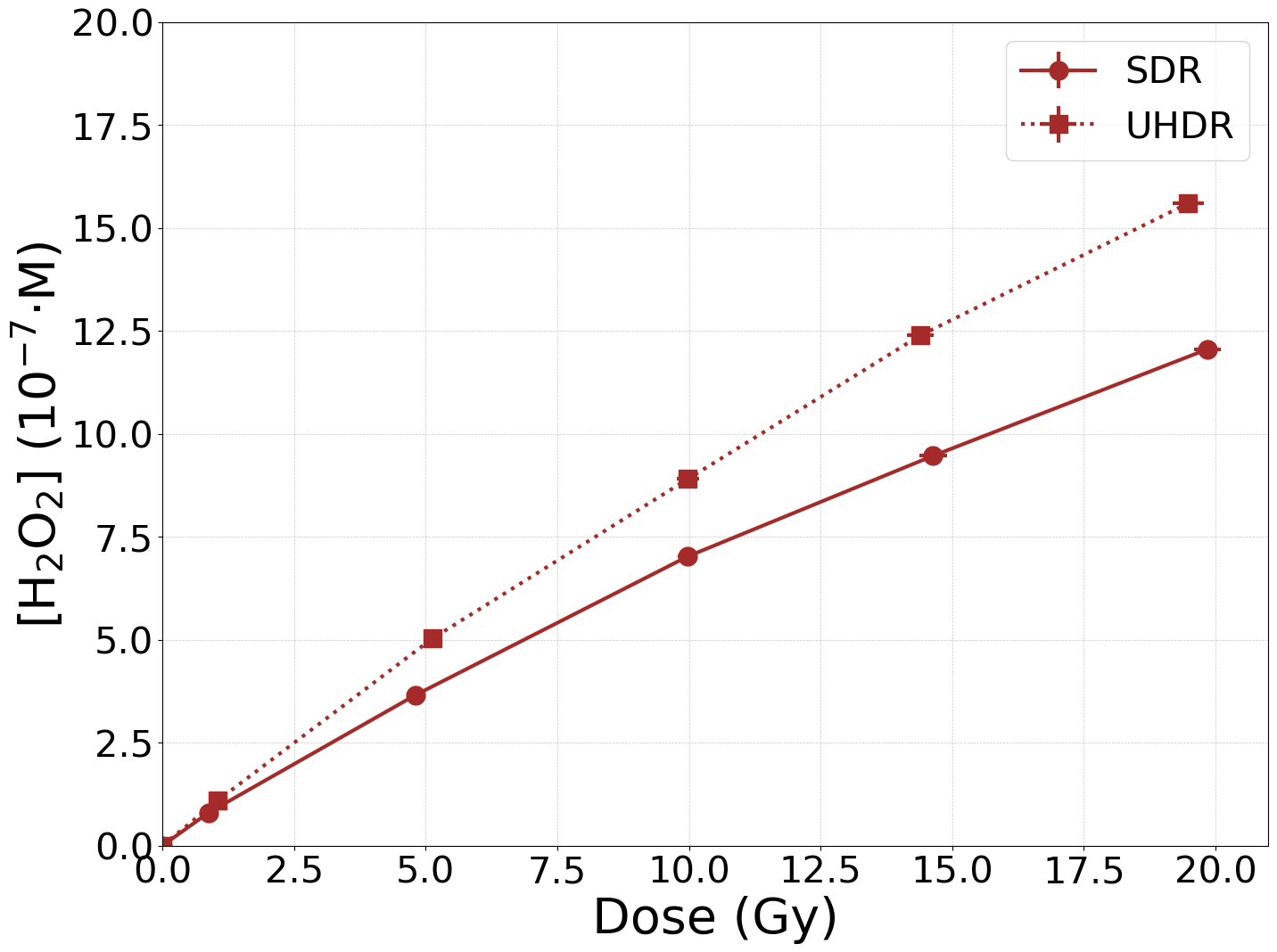}
   \caption{H$_2$O$_2$ concentration against dose following multiple 70 MeV proton pulse irradiations characterized by FWHM = 1 $\muu$s and $\dot{D}_{p}$= 10$^6$ Gy/s. The irradiations are a SDR with a $\dot{D}_{av}$ = 1 Gy/s and a UHDR with $\dot{D}_{av}$ = 10000 Gy/s. Each data point corresponds to the post-irradiation concentration after 1, 5, 10, 15, and 20 pulses.     
    }  %note label inside caption
    \label{Concentration-Dose} 
    \end{center}
\end{figure}

Fig. \ref{Concentration-Dose} displays the post-irradiation H$_2$O$_2$ concentration after delivering 1, 5, 10, 15, and 20 Gy, for SDR and UHDR irradiations. It can be observed that the UHDR irradiation produced more H$_2$O$_2$ than the SDR one: the shorter irradiation time led to a build-up of $^{\cdot}$OH in the system, ultimately enhancing R6, a radical-radical reaction. Additionally, for both cases it is noticeable a root tendency in the H$_2$O$_2$ concentration with the dose. As the dose escalates, the H$_2$O$_2$ accumulates and R5 becomes more predominant, hence H$_2$O$_2$ is consumed.

%%%%%%%%%%%%%%%%%%%%%%%%%%%%%%%%%%%%%%%%%%%%%%%%%%%%%%%%%%%%%%
%%%%%%%%%%%%%%%%%%%%%%%%% Dav-G-values %%%%%%%%%%%%%%%%%%%%%%%%
%%%%%%%%%%%%%%%%%%%%%%%%%%%%%%%%%%%%%%%%%%%%%%%%%%%%%%%%%%%%%%
\begin{figure}[H]
   \begin{center}
   \includegraphics[height=6cm]{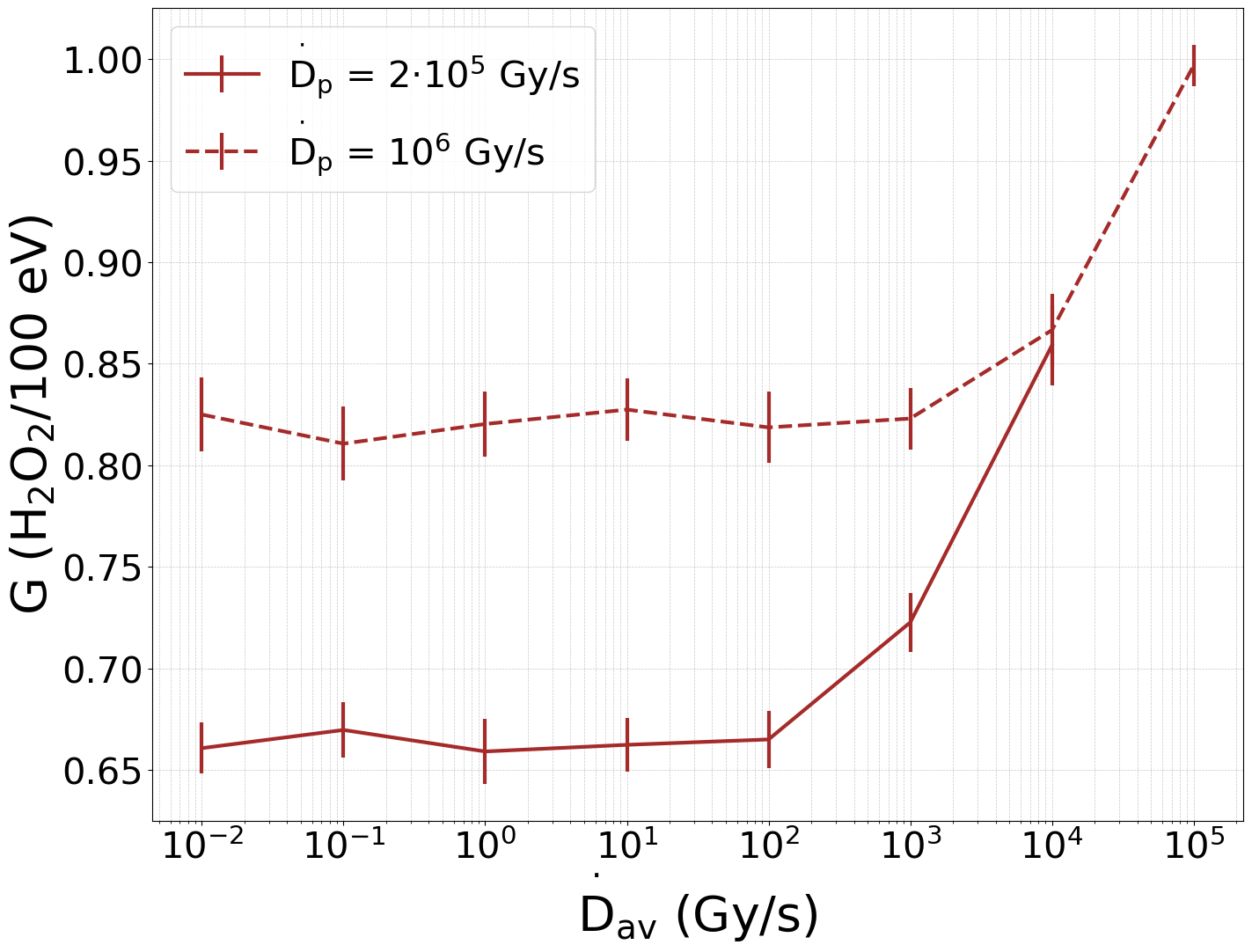}
   \caption{Post-irradiation H$_2$O$_2$ yield as a function of $\dot{D}_{p}$ and $\dot{D}_{av}$ for multiple 70 MeV proton pulse irradiations, characterized by FWHM = 1 $\muu$s. The total dose for every configuration amounted to 20 Gy.}  %note label inside caption
    \label{G-yield_Dav} 
    \end{center}
\end{figure}
Fig. \ref{G-yield_Dav} highlights the H$_2$O$_2$ yield, which was calculated as the slope of the dose-concentration distributions (Fig. \ref{Concentration-Dose}), as a function of $\dot{D}_{av}$ for the two $\dot{D}_{p}$ tested. The yields increase between the SDR $\dot{D}_{av}$ and the maximum UHDR $\dot{D}_{av}$ was 30\% and 20\% for the $\dot{D}_{p}$ of 2·10$^5$ and 10$^6$ Gy/s, respectively. Similarly, the reactive species build-up was amplified with the $\dot{D}_{p}$: the H$_2$O$_2$ production increased by 26\% between the two tested $\dot{D}_{p}$

%%%%%%%%%%%%%%%%%%%%%%%%%%%%%%%%%%%%%%%%%%%%%%%%%%%%%%%%%%%%%%
%%%%%%%%%%%%%%%%%%%%%%%%%%%%%%%%%%%%%%%%%%%%%%%%%%%%%%%%%%%%%%
%%%%%%%%%%%%%%%%%%%%%%%%%%%%%%%%%%%%%%%%%%%%%%%%%%%%%%%%%%%%%%
%%%%%%%%%%%%%%%%%%%%%%% Discussion  %%%%%%%%%%%%%%%%%%%%%%%%%%
%%%%%%%%%%%%%%%%%%%%%%%%%%%%%%%%%%%%%%%%%%%%%%%%%%%%%%%%%%%%%%
%%%%%%%%%%%%%%%%%%%%%%%%%%%%%%%%%%%%%%%%%%%%%%%%%%%%%%%%%%%%%%
%%%%%%%%%%%%%%%%%%%%%%%%%%%%%%%%%%%%%%%%%%%%%%%%%%%%%%%%%%%%%%

\section{Discussion}

%%%%%%%%%%%%%%%%%%%%%%%%%%%%%%%%%%%%%%%%%%%%%%%%%%%%%%%%%%%%%%
% Reactive species build-up promotes radical-radical reactions
%%%%%%%%%%%%%%%%%%%%%%%%%%%%%%%%%%%%%%%%%%%%%%%%%%%%%%%%%%%%%%

\subsection{Reactive species build-up promotes radical-radical reactions}
Simulations of multiple pulse irradiations to pure water showed the significant impact of the reactive species build-up on radical-radical reactions: as the $\dot{D}_{p}$ escalated, the reactive species build-up led to an increase in radical-radical reactions. Under UHDR conditions, the high pulse \textit{f} shortened the intervals between pulses, preventing the achievement of a steady state in the species concentration within the pulses. Consequently, reactive species accumulated, thereby enhancing radical-radical reactions. Below the $\dot{D}_{av}$ threshold (SDR), the same steady state within the pulses was consistently achieved, and the post-irradiation G-values were constant for every $\dot{D}_{av}$. However, above the threshold (UHDR), the steady state was not attained, and reactive species accumulated, radical-radical recombination increased, and leading to G-values dependent on the $\dot{D}_{av}$. The $\dot{D}_{av}$ threshold depended on the $\dot{D}_{p}$ because as the D$_p$ decreases, less chemicals are created, necessitating more time to encounter and neutralize, thereby extending the time to reach the steady state. Therefore, the $\dot{D}_{av}$ threshold is directly proportional to the $\dot{D}_{p}$. The $\dot{D}_{av}$ threshold is also directly proportional to the particle’s LET. As the LET increases, intra-track chemical reactions are enhanced, and the steady state is reached faster. To build-up chemicals with high-LET particles it would be required higher $f$, alternately higher $\dot{D}_{av}$. Future simulations will involve particles with different LET.

%%%%%%%%%%%%%%%%%%%%%%%%%%%%%%%%%%%%%%%%%%%%%%%%%%%%%%%%%%%%%%
%%%%%%% Inter-tracking vs reactive species build-up %%%%%%%%%%
%%%%%%%%%%%%%%%%%%%%%%%%%%%%%%%%%%%%%%%%%%%%%%%%%%%%%%%%%%%%%%

\subsection{Inter-tracking vs. reactive species build-up}
Although both inter-track interactions and reactive species build-up contribute to the enhancement of radical-radical reactions, they are fundamentally distinct mechanisms. Inter-track chemical reactions have a local impact on the G$_0$ due to the entanglement of two or more tracks at times below 1 $\muu$s, an effect within the non-homogeneous stage. In contrast, the reactive species build-up is a macroscopic phenomenon that occurs at later times when the species are homogeneously distributed, a consequence of multiple pulses. The dose rate type that triggers both mechanisms is different; while the inter-track effect is driven by the $\dot{D}_{p}$, the reactive species build-up is influenced by the $\dot{D}_{av}$. A critical distinction between them is the thresholds at which they start to significantly promote radical-radical reactions. As pointed out by Weber et al\cite{weber2022flash} and Thompson et al\cite{thompson2023investigating}, extremely high $\dot{D}_{p}$ are needed to observe inter-track radical reactions, far exceeding the $\dot{D}_{p}$ where the Flash effect has been observed. In the other hand, the results from this study indicate that reactive species build-up promotes radical-radical reactions at $\dot{D}_{av}$ ranging from 10 to 100 Gy/s, aligning with the experimental thresholds for the Flash effect\cite{vozenin2019biological}. Thus, the reactive species build-up is more likely to be a significant promoter of radical-radical reactions under UHDR compared to the inter-track mechanism.

%%%%%%%%%%%%%%%%%%%%%%%%%%%%%%%%%%%%%%%%%%%%%%%%%%%%%%%%%%%%%%
%%%%%%%%%%%%%%%%%%%%%%%%%%%% Enigma %%%%%%%%%%%%%%%%%%%%%%%%%%
%%%%%%%%%%%%%%%%%%%%%%%%%%%%%%%%%%%%%%%%%%%%%%%%%%%%%%%%%%%%%%

\subsection{The hydrogen peroxide enigma: experiments vs. models}
In this work, above the $\dot{D}_{av}$ threshold it was observed an enhancement of R6, which resulted in a higher H$_2$O$_2$ yield. This is consistent with the traditional pulse radiolysis literature, which anticipates an increase in the H$_2$O$_2$ yield as the dose rate escalates\cite{wardman2020radiotherapy}. However, the experimental data from recent years points in the opposite direction, as they consistently show a reduction in the H$_2$O$_2$ under UHDR.  This paradox is not well understood. The H$_2$O$_2$ yield ratio between SDR and UHDR irradiations for this work and for various experimental trials is presented in Tabl \ref{tab:H2O2ResultsComparison}. Note that these results were obtained in in aerated water mediums, while the simulations from this work were performed in pure water. Investigating the methods used to measure H$_2$O$_2$ might provide insights. These methods were the Amplex Red assay kit\cite{montay2019long, kacem2022comparing, sunnerberg2023mean} and the Ghormley triiodide method\cite{blain2022proton}. It could be hypothesized that the counterintuitive reduction of H$_2$O$_2$ under UHDR could potentially be caused by an unknown effect of radiation on these assays. On the other hand, there is a significant need to improve the MCTS algorithms. To unravel the H$_2$O$_2$ mystery, more complex chemistry lists that include oxygen and pH reactions should be considered. 
\vspace{0.5cm}
\begin{table}[H]
\centering
\small
%\scriptsize
\setlength{\tabcolsep}{2pt} % Adjust the space between columns
\renewcommand{\arraystretch}{1.5} 
\begin{tabular}{c|c|ccc|ccc|c|}
\cline{2-9}
  
    &  \multirow{2}{*}{\textbf{Particle}} & \multicolumn{3}{c|}{\textbf{SDR}} &  \multicolumn{3}{c|}{\textbf{UHDR}} &  \multirow{2}{*}{\textbf{G Ratio (\%)}}\\
    \cline{3-8}
    
    &  & $\dot{D}_p$ & $\dot{D}_{av}$ & G & $\dot{D}_p$ & $\dot{D}_{av}$ & G& \\
    \hline
    
    \multicolumn{1}{|c|}{This work} & 70 MeV proton & 2$\cdot10^5$ & 0.01 & 0.66 &  2$\cdot10^5$ & 10000 & 0.86 & 30.\\
    \cline{3-9}
    \multicolumn{1}{|c|}{} & & $10^6$ & 0.01 & 0.83 &  $10^6$ & 100000 & 1.00 & 20.\\
    \hline
    
    \multicolumn{1}{|c|}{Montay-Gruel \textit{et al.}\cite{montay2019long}} & 6 MeV electron & 3.0$\cdot10^4$ & 0.29 & 1.42 & 2.8$\cdot10^6$ & 500 & 1.20 & -16.\\
    \hline

    \multicolumn{1}{|c|}{Kacem \textit{et al.}\cite{kacem2022comparing}} & 5.5 MeV electron & $10^4$ & 0.1 & 2.81 & 5.56$\cdot10^5$ & 100 & 2.33 & -17. \\
    \cline{2-9}
    \multicolumn{1}{|c|}{} & 235 MeV proton & - & 0.9 & 2.33 & - & 1260 & 1.92 & -18. \\
    \hline

    \multicolumn{1}{|c|}{Blain \textit{et al.}\cite{blain2022proton}} & 68 MeV proton & - & 0.2 & 0.98 & - & 60000 & 0.58 & -40\\
    \hline

    \multicolumn{1}{|c|}{Sunnerberg \textit{et al.}\cite{sunnerberg2023mean}} & 10 MeV electron & 10$^2$ & 0.14 & 2.05 & 10$^6$ & 1500 & 0.65 & -68.\\

\hline
\end{tabular}
\caption{Comparison of G(H$_2$O$_2$/100 eV) between SDR and UHDR irradiations, incorporating the simulations results from this work, and experimental measurements. The data is presented indicating the $\dot{D}_p$ (Gy/s) and $\dot{D}_{av}$ (Gy/s) settings. Note that for Kacem \textit{et al.}\cite{kacem2022comparing}, and Blain \textit{et al.}\cite{blain2022proton} used continuous proton beams and no $\dot{D}_p$ is shown.}
\label{tab:H2O2ResultsComparison}
\end{table}

\section{Conclusions}
This investigation demonstrated that reactive species build-up significantly influences the enhancement of radical-radical reactions, a pivotal discussion in the Flash effect\cite{jansen2022changes}. We demonstrated that it is a far more likely radical-radical reaction promoter than the inter-track mechanism. Under UHDR conditions, the shortened beam-off time prevents the steady-state in the species concentration within the pulses, resulting in the accumulation of reactive species. This phenomenon promotes radical-radical reactions at dose rates thresholds similar to those at which the Flash effect was observed \cite{vozenin2019biological}. In mice and zebrafish, it was observed for $\dot{D}_{av}\geq40$ Gy/s and $\dot{D}_{p}=1.8\cdot10^5$ Gy/s. For pigs and cats for $\dot{D}_{av}>100$ Gy/s and $\dot{D}_{p}=10^6$ Gy/s. These thresholds algin with those from our simulations, suggesting a correlation between the Flash effect and chemistry. Although current models predict an opposing trend in the H$_2$O$_2$ yield to the experimental results, the reactive species build-up still provides an explanation for the chemical dose rate dependencies. This discrepancy likely stem from the models not being fully calibrated to represent all the experimental nuances. Once resolved, it is anticipated that the chemical kinetics and dose rate thresholds we discuss in this work prevail, reinforcing the importance of the reactive species build-up in understanding dose rate effects. The advancements on gMicroMC we developed in this work sets the foundations for future models that will help understand the underlying mechanisms of the Flash effect. The ultimate goal is to extend the model for biological systems. Although water is commonly used as a soft tissue equivalent for dosimetric calculations in standard radiotherapy, the chemical effect of dose rate in water may not directly correlate to the chemistry of biological systems. Nonetheless, developing water models that contribute to the understanding of the role of dose rate in water’s chemistry establish a foundation for constructing more accurate biological models. This approach, akin to “putting the horse before the cart”, where the “horse” represents water’s chemistry and the “cart” biological systems.

%%%%%%%%%%%%%%%%%%%%%%%%%%%%%%%%%%%%%%%%%%%%%%%%%%%%%%%%%%%%%%
%%%%%%%%%%%%%%%%%%%%%%%%%%%%%%%%%%%%%%%%%%%%%%%%%%%%%%%%%%%%%%
%%%%%%%%%%%%%%%%%%%%%%%%%%%%%%%%%%%%%%%%%%%%%%%%%%%%%%%%%%%%%%
%%%%%%%%%%%%%%%%%%%% Acknowledgments  %%%%%%%%%%%%%%%%%%%%%%%%
%%%%%%%%%%%%%%%%%%%%%%%%%%%%%%%%%%%%%%%%%%%%%%%%%%%%%%%%%%%%%%
%%%%%%%%%%%%%%%%%%%%%%%%%%%%%%%%%%%%%%%%%%%%%%%%%%%%%%%%%%%%%%
%%%%%%%%%%%%%%%%%%%%%%%%%%%%%%%%%%%%%%%%%%%%%%%%%%%%%%%%%%%%%%

\section*{Acknowledgments}
This work was funded through a PhD fellowship from the PT-CERN project, a collaboration between the Foundation for Science and Technology, an organization within the Ministry of Science, Technology and Higher Education in Portugal and the Laboratory of Instrumentation and Experimental Particles Physics in Lisbon, Portugal \href{https://doi.org/10.54499/SFRH/BD/151007/2021}{\textcolor{blue}{https://doi.org/10.54499/SFRH/BD/151007/2021}}. This work was also supported by the Deutsche Krebshilfe DKH grant with reference numbers 70115332, 70115445, and entitled “Dosisleistungsabhangige Anderung des Sauerstoffpartlaldrucks wahrend FLASH-Bestrahlung und deren Einfluss auf die strahlenbiologische Wirkung in Zebrafisch Embryonen”. Yujie Chi is partially supported by the National Cancer Institute grant 1R15CA256668-01A1.

%%%%%%%%%%%%%%%%%%%%%%%%%%%%%%%%%%%%%%%%%%%%%%%%%%%%%%%%%%%%%%
%%%%%%%%%%%%%%%%%%%%%%%%%%%%%%%%%%%%%%%%%%%%%%%%%%%%%%%%%%%%%%
%%%%%%%%%%%%%%%%%%%%%%%%%%%%%%%%%%%%%%%%%%%%%%%%%%%%%%%%%%%%%%
%%%%%%%%%%%%%%%%% Conflict of interest  %%%%%%%%%%%%%%%%%%%%%%
%%%%%%%%%%%%%%%%%%%%%%%%%%%%%%%%%%%%%%%%%%%%%%%%%%%%%%%%%%%%%%
%%%%%%%%%%%%%%%%%%%%%%%%%%%%%%%%%%%%%%%%%%%%%%%%%%%%%%%%%%%%%%
%%%%%%%%%%%%%%%%%%%%%%%%%%%%%%%%%%%%%%%%%%%%%%%%%%%%%%%%%%%%%%

\section*{Conflict of interest}
The authors declare no conflict of interest.

%%%%%%%%%%%%%%%%%%%%%%%%%%%%%%%%%%%%%%%%%%%%%%%%%%%%%%%%%%%%%%
%%%%%%%%%%%%%%%%%%%%%%%%%%%%%%%%%%%%%%%%%%%%%%%%%%%%%%%%%%%%%%
%%%%%%%%%%%%%%%%%%%%%%%%%%%%%%%%%%%%%%%%%%%%%%%%%%%%%%%%%%%%%%
%%%%%%%%%%%%%%%% Supplementary Material  %%%%%%%%%%%%%%%%%%%%%
%%%%%%%%%%%%%%%%%%%%%%%%%%%%%%%%%%%%%%%%%%%%%%%%%%%%%%%%%%%%%%
%%%%%%%%%%%%%%%%%%%%%%%%%%%%%%%%%%%%%%%%%%%%%%%%%%%%%%%%%%%%%%
%%%%%%%%%%%%%%%%%%%%%%%%%%%%%%%%%%%%%%%%%%%%%%%%%%%%%%%%%%%%%%

\section*{Supplementary material}

\setcounter{table}{0}
\setcounter{figure}{0}

\renewcommand{\thetable}{S\arabic{table}}
\renewcommand{\thefigure}{S\arabic{figure}}

\begin{table}[H]
\small
\centering
\begin{tabular}{cc|c|c|}
\cline{3-4}
                                                                 &                                           & \textbf{Chanel}                            & \textbf{Probability} \\ \hline
\multicolumn{1}{|c|}{\textbf{Ionization}}                        & H$_2$O$^+$                                & H$^{+}$ + $^{\cdot}$OH                     & 100                  \\ \hline
\multicolumn{1}{|c|}{\multirow{9}{*}{\textbf{Excitation}}}       & \multirow{2}{*}{A$^1$B$_1$}               & H$^{\cdot}$ + $^{\cdot}$OH                 & 65                   \\ \cline{3-4} 
\multicolumn{1}{|c|}{}                                           &                                           & H$_2$O                                     & 35                   \\ \cline{2-4} 
\multicolumn{1}{|c|}{}                                           & \multirow{5}{*}{B$^1$A$_1$}               & H$^{+}$ + $^{\cdot}$OH + e$_{\text{aq}}^-$ & 50                   \\ \cline{3-4} 
\multicolumn{1}{|c|}{}                                           &                                           & H$^{\cdot}$ + $^{\cdot}$OH                 & 25.35                \\ \cline{3-4} 
\multicolumn{1}{|c|}{}                                           &                                           & H$_2$ + 2$^{\cdot}$OH                      & 3.25                 \\ \cline{3-4} 
\multicolumn{1}{|c|}{}                                           &                                           & 2H$^{\cdot}$ + O($^3$P)                    & 3.9                  \\ \cline{3-4} 
\multicolumn{1}{|c|}{}                                           &                                           & H$_2$O                                     & 17.5                 \\ \cline{2-4} 
\multicolumn{1}{|c|}{}                                           & \multirow{2}{*}{Rydberg, Diffusion Bands} & H$^{+}$ + $^{\cdot}$OH + e$_{\text{aq}}^-$ & 50                   \\ \cline{3-4} 
\multicolumn{1}{|c|}{}                                           &                                           & H$_2$O                                     & 50                   \\ \hline
\multicolumn{1}{|c|}{\multirow{5}{*}{\textbf{Electron Capture}}} & Electron Attachment                       & $^{\cdot}$OH + OH$^{-}$ + H$_2$            & 100                  \\ \cline{2-4} 
\multicolumn{1}{|c|}{}                                           & \multirow{4}{*}{Electron Hole}            & H$^{\cdot}$ + $^{\cdot}$OH                 & 35.75                \\ \cline{3-4} 
\multicolumn{1}{|c|}{}                                           &                                           & H$_2$ + 2$^{\cdot}$OH                      & 13.65                \\ \cline{3-4} 
\multicolumn{1}{|c|}{}                                           &                                           & 2H$^{\cdot}$ + O($^3$P)                    & 15.6                 \\ \cline{3-4} 
\multicolumn{1}{|c|}{}                                           &                                           & H$_2$O                                     & 35                   \\ \hline
\multicolumn{1}{|c|}{\textbf{Electron Solvation}}                & Hydratation                               & e$_{\text{aq}}^-$                          & 100                  \\ \hline
\end{tabular}
\caption{ Branching ratios and associated probability for each channel included in gMicroMC collected from Shin et al.\cite{shin2021geant4}.}
\label{branchingratios}
\end{table}
\vspace{0.5cm}

\begin{table}[H]
\centering
\begin{tabular}{|c|c|}
\hline
\textbf{Chemical Species}  & \textbf{$\mathcal{D}$ (10$^9\cdot$nm$^2$s$^{-1}$)} \\ \hline
e$_{\text{aq}}^-$ & 4.82                          \\ \hline
$^{\cdot}$OH      & 2.2                           \\ \hline
H$^{\cdot}$       & 7.0                           \\ \hline
H$^{+}$           & 9.3                           \\ \hline
H$_2$             & 4.8                           \\ \hline
OH$^{-}$          & 5.3                           \\ \hline
H$_2$O$_2$        & 2.3                           \\ \hline
\end{tabular}

\caption{Chemical species and their respective $\mathcal{D}$ included in gMicroMC collected from Elliot et al.\cite{elliot1994rate}.}
\label{DiffusionContinued}
\end{table}

%%%%%%%%%%%%%%%%%%%%%%%%%%%%%%%%%%%%%%%%%%%%%%%%%%%%%%%%%%%%%%
%%%%%%%%%%%%%%%%%%%%%%%% Flow chart %%%%%%%%%%%%%%%%%%%%%%%%%%
%%%%%%%%%%%%%%%%%%%%%%%%%%%%%%%%%%%%%%%%%%%%%%%%%%%%%%%%%%%%%%
\begin{figure}[ht]
   \begin{center}
   \includegraphics[height=12cm]{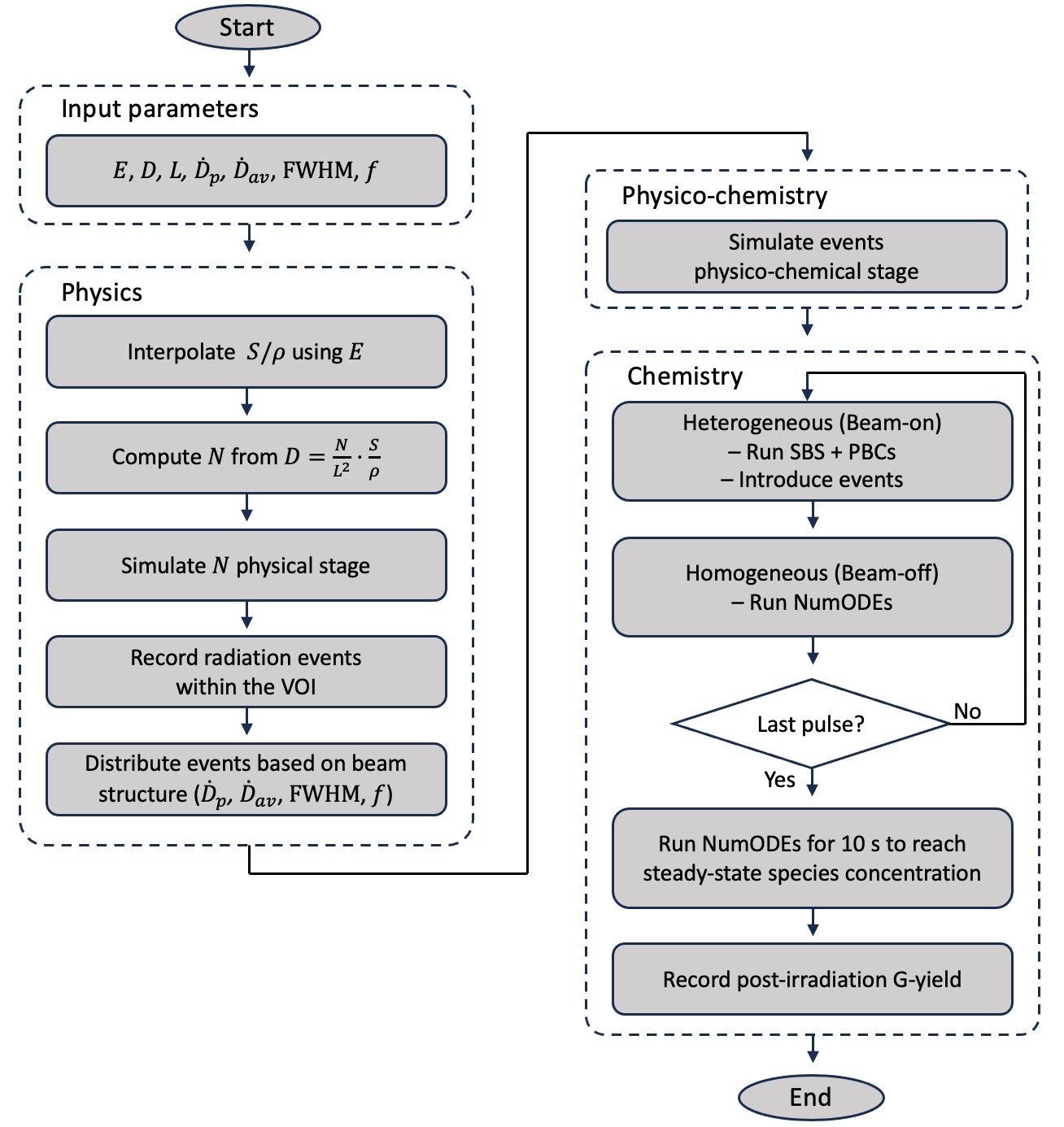}
   \caption{Flow chart depicting the gMicroMC extension for multiple pulse irradiations.   
    }  %note label inside caption
    \label{Flowchart} 
    \end{center}
\end{figure}

\vspace{0.5cm}
\begin{figure}[H]
   \begin{center}
   \includegraphics[height=6cm]{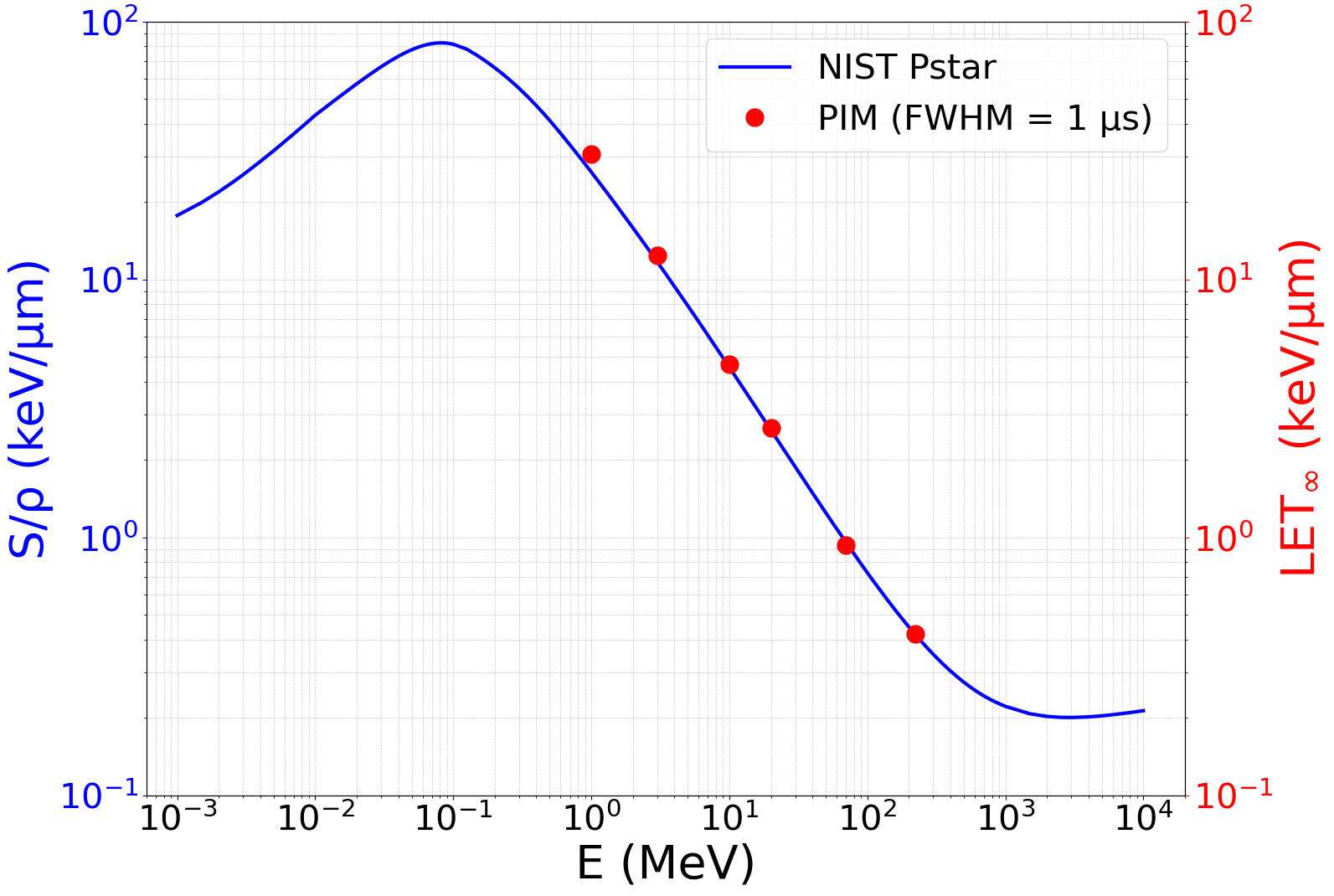}
   \caption{LET$_\infty$ within a VOI characterized by $L$ = 5 $\mu$m for single pulse radiolysis characterized by FWHM = 1 $\mu$s and $\dot{D}_{p}$ = 10$^6$ Gy/s. It is shown the $S/\rho$ obtained from NIST Pstar\cite{berger1998stopping} for the corresponding $E$.   
    }  %note label inside caption
    \label{Nist} 
    \end{center}
\end{figure}

\newpage
\subsection*{Reaction radius calculation for self-radical recombination reactions}

The mathematical analysis for the calculation of the reaction radius ($R$)  for diffusion controlled reactions is presented below, utilizing the Smoluchowski phenomenological approach\cite{bamford1985diffusion}. Specifically, it is adapted for self-radical recombination reactions, such as the \ce{$^{\cdot}$OH} recombination:

\begin{equation}
    \ce{$^{\cdot}$OH + $^{\cdot}$OH -> H2O2}
    \label{eq:OHrecombination}
\end{equation}

Assuming two species A and B (Eq. \ref{eq:simplereaction}), the reaction rate in terms of the species concentration is given by the rate equation (Eq. \ref{eq:rateequation}). For the self-radical recombination reactions a factor 2 has to be considered, as shown in Eq. \ref{eq:rateequation2}.

\begin{equation}
    \ce{A + B -> Products}
    \label{eq:simplereaction}
\end{equation}

\begin{equation}
    -\frac{d[\text{A}]}{dt}=-\frac{d[\text{B}]}{dt}=k_{obs}\,[\text{A}]\,[\text{B}]
    \label{eq:rateequation}
\end{equation}

\begin{equation}
    -\frac{d[^{\cdot}\text{OH}]}{dt}=2\,k_{obs}\,[^{\cdot}\text{OH}]^2
    \label{eq:rateequation2}
\end{equation}

The rate at which the reaction takes place is denoted as $k_{obs}$. It is measured experimentally and it has two components: the intrinsic rate $k_{react}$ and the diffusion rate $k_{diff}$. The three are related by the Noyes equation:

\begin{equation}
    \frac{1}{k_{\text{obs}}} = \frac{1}{k_{\text{react}}} + \frac{1}{k_{\text{diff}}}
    \label{eq:noyes}
\end{equation}

For diffusion-controlled reactions, the species react immediately upon encounter when their mutual distance $r$ is smaller than the reaction radius $R$. For these type of reactions, $k_{obs}=k_{diff}$. Following, it is described the procedure to link $k_{diff}$ and $R$. Initially, consider the probability density function (Eq. \ref{eq:PDF0}) of B species diffusing towards A species, where $\text{[B]}_0$ is the initial concentration of B and $\mathcal{D}_{\text{AB}}=\mathcal{D}_{\text{A}}+\mathcal{D}_{\text{B}}$\cite{bamford1985diffusion}. The initial conditions specify that $p(r, 0)$ = 1 if $r>R$ and $p(r, 0)$ = 0 if $r \leq R$, indicating that initially A and B have not yet reacted. The boundary conditions are defined such that $p(r \rightarrow \infty, t)$ = 1 and  $p(r \leq R, t)$ = 0, implying that A and B react if they encounter at a distance below $R$.

\begin{equation}
    p(r, t) = \frac{\text{[B](r)}}{\text{[B]}_0} = 1-\frac{\mathcal{R}}{r}\, \text{erfc} \left\{
        -\frac{r-\mathcal{R}}{\sqrt{4\,\mathcal{D}_{\text{AB}} \,t}}
    \right\}
    \label{eq:PDF0}
    \end{equation}

The current $I$ is the number of B reactants diffusing towards A species through an area $4\pi r^2$ per unit of time. It is used to link the $p(r,t)$ with the reaction law and it can be expressed in terms of $p(r,t)$ and the flux $J$ (Fick's first law) as follows:

\begin{equation}
    I=4\pi r^2 J = 4\pi r^2 \mathcal{D}_{AB} \frac{\partial \text{[B]}}{\partial r} \bigg|_{\mathcal{R}}=4\pi r^2 \mathcal{D}_{AB} \frac{\partial p(r, t)}{\partial r} \bigg|_{\mathcal{R}}\text{[B]}_0
    \label{eq:current0}
\end{equation}

\begin{equation}
    I=4\pi \mathcal{R} \,\mathcal{D}_{AB} \left(1 + \frac{\mathcal{R}}{\sqrt{\pi \, \mathcal{D}_{AB} \, t}}\right) [\text{B}]_0
    \label{eq:current1}
\end{equation}

Alternatively, $I$ can be expressed in terms of the reaction law (Eq. \ref{eq:current2}). For self-radical recombination reactions the factor 2 must be considered, as displayed in Eq. \ref{eq:current2self}.

\begin{equation}
    I=-\frac{1}{[\text{A}]}\frac{d[\text{A}]}{dt}=k_{diff} \,[\text{B}]
    \label{eq:current2}
\end{equation}

\begin{equation}
    I=-\frac{1}{[^{\cdot}\text{OH}]}\frac{d[^{\cdot}\text{OH}]}{dt}=2\,k_{diff} \,[^{\cdot}\text{OH}]
    \label{eq:current2self}
\end{equation}

By comparing Eq. \ref{eq:current1} and Eq. \ref{eq:current2}, $k_{diff}$ can be obtained by multiplying by $N_a$ (Eq. \ref{eq:kdiff1}). For self-radical recombination reactions the factor 2 is considered, as displayed in Eq. \ref{eq:kdiff2}.

\begin{equation}
    k_{diff}(t) = 4 \pi \, \mathcal{R}\, \mathcal{D}_{\text{AB}} \, \left(1 + \frac{\mathcal{R}}{\sqrt{\pi \, \mathcal{D}_{\text{AB}} \, t}}\right) N_a
    \label{eq:kdiff1}
\end{equation}

\begin{equation}
   2\, k_{diff}(t) = 4 \pi \, \mathcal{R} \, \mathcal{D}_{^{\cdot}\text{OH}^{\cdot}\text{OH}} \, \left(1 + \frac{\mathcal{R}}{\sqrt{\pi \, \mathcal{D}_{^{\cdot}\text{OH}^{\cdot}\text{OH}} \, t}}\right) N_a
   \label{eq:kdiff2}
\end{equation}
\newpage
\section*{References}
\vspace{-1cm}
\setlength{\bibsep}{0pt plus 0.3ex}
\bibliographystyle{./medphy.bst}
\bibliography{references}

\end{document}